\documentclass{aa}  
\usepackage{comment}
\usepackage{bm}
\usepackage{soul}
\usepackage{natbib}
\usepackage{placeins}

\def \cii {C{\sc ii}}
\def \ci {C{\sc i}}
\def \Htwo {$\rm H_{2}$}

\def \mum  {$\rm \mu$m}
\def \LIR   {$\rm L_{IR}$}
\def \msun {M_\odot}
\def \mci {$ \rm M(H_2)^{[CI]}$}
\def \mco {$\rm M(H_2)^{CO}$}

\newcommand{\mycolor}[1]{{\color{black}#1}}
\newcommand{\mysecondcolor}[1]{{\color{black}#1}}
\newcommand{\mythirdcolor}[1]{{\color{black}#1}}

\usepackage{txfonts}
\usepackage{graphicx}
\usepackage{amsmath}	
\usepackage{booktabs}
\usepackage{xcolor}
\usepackage{hyperref}
\hypersetup{
   colorlinks=true,
    linkcolor=blue,
    urlcolor=blue,
   citecolor=blue
    }
\begin{document}

\title{Tight correlation of star formation with [\ci] and CO lines across cosmic time}
\titlerunning{SFR tracers}


   \author{Theodoros Topkaras$^1$,\thanks{E-mail: topkaras@ph1.uni-koeln.de}
            Thomas G. Bisbas$^2$,\thanks{E-mail: tbisbas@zhejianglab.com}
            Zhi-Yu Zhang$^{3,4}$, and
            V. Ossenkopf-Okada$^1$
            }
    \institute{I. Physikalisches Institut, Universität zu Köln, Zülpicher Str. 77, 50937 Köln, Germany 
    \and Research Center for Astronomical Computing, Zhejiang Lab, Hangzhou 311100, China
    \and School of Astronomy and Space Science, Nanjing University, Nanjing, China
    \and Key Laboratory of Modern Astronomy and Astrophysics, Nanjing University, Ministry of Education, Nanjing, China\\}

   \date{Accepted XXX. Received YYY; in original form ZZZ}

  \abstract
  {
  Cold molecular gas tracers, such as
  \ci\ and CO lines, have been widely used to infer specific characteristics of the ISM and
  to derive star-formation relations among galaxies.}
  {However, there is still a lack of systematic studies of the star-formation
  scaling relation of CO and [\ci] lines across cosmic time on multiple physical
  scales.} 
  { \mycolor{We used observations of the ground state transitions of [\ci], CO, and [\cii], 
  for 885 sources collected
  from the literature, to infer possible correlations between line luminosities of $\rm 
  L^{'}_{[CI](1-0)}, \rm L^{'}_{CO(1-0)}$, and $\rm L^{'}_{[CII]}$ with star
  formation rates (SFR).} With linear regression,  we fit the relations between
  SFR and molecular mass derived from  CO, \ci, and \cii\ lines.}
{ \mysecondcolor{The relation between [\ci] and CO based total molecular masses is weakly superlinear. Nevertheless, they can be calibrated against each other. For $\rm \alpha_{CO} = 0.8$ and $4.0\ \rm {M}_{\odot}\,({K}\,{km}\,{s}^{-1}\,{pc}^2)^{-1}$ we derive $\alpha_{\rm [CI]} = 3.9$ and $\sim$$17\ \rm {M}_{\odot}\,({K}\,{km}\,{s}^{-1}\,{pc}^2)^{-1}$ , respectively. Using the \emph{lmfit} package,} we derived relation slopes of SFR--$\rm L^{'}_{[CI](1-0)}$, SFR--$\rm L^{'}_{CO(1-0)}$, and SFR--$\rm L^{'}_{[CII](1-0)}$ to be $\rm \beta$ = 1.06 $\pm$ 0.02, 1.24 $\pm$ 0.02, and 0.74 $\pm$ 0.02, respectively. With a Bayesian-inference \emph{linmix} method, we find consistent results.}
{Our relations for [\ci](1-0) and CO(1-0) indicate that they trace
 similar \mysecondcolor{molecular} gas contents, across different redshifts and different types of
 galaxies. This suggests that these correlations do not have strong evolution
 with cosmic time.}


   \maketitle

%

\section{Introduction}\label{sec:into}

By studying the star formation rate (SFR) across different cosmic times and its
correlation with other measurable quantities
\citep{elbaz2011goods,kennicutt2012star,jo2021star}, the subsequent
evolution and formation of galaxies can be
unraveled \citep[see][for a
review]{McKee07}. Some of the first quantitative SFR estimates were derived based on
an evolutionary synthesis of galaxy colors
\citep{tinsley1968evolution,tinsley1972galactic}, followed by
\citet{kennicutt1998star} who studied the SFR using measurements in far-infrared
(FIR), ultraviolet (UV),  and nebular recombination lines. The aforementioned works show that the \mycolor{cosmic SFR} follows a
hierarchy of physical processes. Objects spanning from Mega-parsec (Mpc) to
kilo-parsec (kpc) scales (e.g., the intergalactic medium `IGM' or spiral arms)
collapse into smaller-scale structures leading to molecular cloud formation 
\mycolor{\citep{dobbs2014formation,inutsuka2015formation}}. 
The latter may collapse even further and fragment to form dense clumps of sub-pc
scales, and eventually progenitors of cores, and planetary systems 
\mycolor{\citep{larson1973processes,boss2001formation,guszejnov2016star}}.

Previous studies estimated the SFR using extrapolated mid-infrared (mid-IR) or submillimeter observations \citep{calzetti2007calibration, kennicutt2012star, simpson2015scuba, da2015alma}. More recent work, however, has shifted toward using the total infrared luminosity ($\rm L_{IR}$, integrated over 8–1000 $\rm \mu$m), which captures the emission from dust-obscured, newly formed stars \citep{stacey2021rocky, montoya2023sensitive}. Although luminous infrared galaxies (LIRGs) were initially thought to be the dominant contributors to the cosmic SFR density at redshifts around $\textit{z} \sim 1$ \citep{chary2001interpreting, le2005infrared, magnelli20090}, later studies showed that more luminous sources, named ultra-luminous infrared galaxies (ULIRGs), with \LIR$>10^{12}$\, L$_\odot$, also contributed significantly at higher redshifts ($\textit{z} \gtrsim 2$) \citep{daddi2007multiwavelength, magnelli20090, magnelli2011evolution, Wang2019Natur}. \citet{jo2021star} examined the evolution of the cosmic SFR density across a wide redshift range ($0 < \textit{z} < 6$), considering both major and minor contributors to star formation. Similarly, \citet{elbaz2011goods}, using far-infrared data from the \textit{Herschel} Space Observatory (GOODS-\textit{Herschel} key program), explored the SFR density and mid-IR continuum sizes. Both studies found that the SFR and $\rm L_{IR}$ follow a log-normal distribution with redshift; rising steeply at $\textit{z} < 2$ and plateauing at $\textit{z} > 2$. This reveals the tight correlation between SFR and $\rm L_{IR}$ (see Fig. 4 in \citealt{elbaz2011goods} and Fig. 2 in \citealt{jo2021star}).

\mycolor{The two most abundant cold molecular gas tracers are the ground state transition 
of $^{12}$CO($J$=1-0) at 115.27 GHz (hereafter CO(1-0)), and the ground state
transition of [\ci] $({}^{3}P_1 - {}^{3}P_0)$ at 492.16\,GHz (hereafter [\ci](1-0)}\footnote{The [\ci] notation denotes the line's emission, whereas the
C notation denotes the actual species and/or its abundance. [\ci] only means
it is a forbidden line. It has no meaning of abundance.}). Their correlation 
with SFR, namely, the SFR--$\rm L^{'}_{CO(1-0)}$,
and the SFR--$\rm L^{'}_{[CI](1-0)}$ \mysecondcolor{relations} \mycolor{provide a comparison} to the integrated Schmidt-Kennicutt (S-K) law \citep[][see also \citealt{kennicutt2012star}]{Schmidt1959}. The S-K law empirically
links the surface density of cold gas to that of the star formation rate (SFR)
expressed as $\Sigma_{\rm SFR}$ $\propto$ $\Sigma_{\rm gas}$. This fundamental
scaling relation \citep{Schmidt1959,kennicutt1998global} has been found to hold across a wide range of
conditions, spanning several orders of magnitude both in $\Sigma_{\rm SFR}$ and
$\Sigma_{\rm gas}$. Notably, studies like \citet{gao2004star, Gao2004ApJS,
wu2005connecting,zhang2014dense,zhou2022dense} have reaffirmed that dense gas is more tightly related to star formation, compared to total \mycolor{neutral} gas. In this work, we examine the relationship between SFR and the luminosities of CO(1–0) and [\ci][1-0]. Specifically, we test the SFR--$\rm L^{'}_{CO(1-0)}$ and SFR–-$\rm L^{'}_{[CI](1-0)}$ \mysecondcolor{relations} to evaluate the potential of these lines as reliable tracers of star formation.

CO molecular lines, particularly low-J transitions like CO(1–0) and CO(2–1), are widely used to trace the cold H$_2$ content of the interstellar medium (ISM), providing a robust means of estimating the total molecular gas mass \citep{bothwell2013survey,boogaard2020alma,birkin2021alma,montoya2023sensitive}. This is typically done by applying a CO-to-H$_2$ conversion factor, $\rm \alpha_{CO}$, to the observed CO(1–0) line luminosity ($\rm L^{'}_{CO(1-0)}$). A commonly adopted value for $\rm \alpha_{CO}$ in the molecular ISM of disk galaxies (e.g., the Milky Way) is $\rm 4.3\,{M}_{\odot}\,({K}\,{km}\,{s}^{-1}\,{pc}^2)^{-1}$ \citep{bolatto2013co}. In contrast, significantly lower values of $\rm \alpha_{CO} \sim 0.8$–$1.0$ are typically used for more dynamically active systems, such as starbursts or galaxies hosting an Active Galactic Nuclei (AGN), where elevated star formation rates and feedback processes may lead to higher excitation conditions \citep{downes1998rotating}. However, the universality of such a large difference in the value of the $\rm \alpha_{CO}$ factor has been increasingly questioned \citep[see recent works of][]{harrington2021turbulent,dunne2022dust,berta2023}. \mysecondcolor{This contrasts with earlier results from \citet{downes1998rotating}, who based their conclusions on a small sample of low-J CO transitions in (U)LIRGs, potentially missing the warm, diffuse, and dense molecular gas present in local IR-luminous star-forming galaxies.} This conversion factor has also been the subject of extensive investigation through both observational studies \citealt{magdis2011goods,jiao2021carbon,teng2022molecular} and numerical models \citep{feldmann2012x,gong2020environmental,bisbas2021photodissociation}, as it is known to depend strongly on environmental conditions in the ISM, such as FUV radiation field strength, and cosmic ray ionization rate \citep[see also review by][]{bolatto2013co}. In addition, it is known to be sensitive to metallicity
\citep{genzel2012metallicity,shi2016carbon,schruba2017physical}, and appears to
have smaller values at galactic centers \citep{bolatto2013co}. \citet{sandstrom2013co}, for instance, showed that $\rm \alpha_{CO}$ values in galactic centers can be up to a factor of $\sim$2 lower than galaxy-wide averages, though the weak correlation with metallicity in their sample suggests other ISM properties (e.g., high gas temperatures) play a significant role.

The fine structure emission of atomic carbon, [\ci](1-0), can also be
used to infer the ISM characteristics
\citep{papadopoulos2004c,walter2011survey,valentino2018survey,valentino2020properties,harrington2021turbulent}.
A significant difference compared to CO is that [\ci](1-0) is mostly optically thin
\citep{ikeda2002distribution,perez2015detection,harrington2021turbulent}, while
CO transitions are typically optically thick, with $\tau \, \geq$ 10 at
$\rm \Sigma_{SFR}$ > 1 $\rm Myr^{-1} \,kpc^{-2}$ \citep{narayanan2014theory}.
In addition, the increasing contribution of the cosmic microwave background (CMB)
radiation with redshift, affects the [\ci] emission to a lesser degree
\citep{zhang2016gone}.
Furthermore, \citet{bisbas2015effective} suggested that elevated cosmic-ray (CR)
ionization rates ($\rm \zeta_{\rm CR}$) can lead to the destruction of CO
molecules while the gas remains H$_2$-rich. 
\mythirdcolor{High star-forming environments are expected to have elevated
$\rm \zeta_{CR}$ \citep{luo2023dependence}, suggesting that [\ci](1-0) could potentially
measure the molecular gas content more accurately as opposed
to low-J CO transitions, in those environments. Consequently,
regions with enhanced SFR activity may eventually become deficient in CO emission, while simultaneously being amplified
in [\ci] and possibly [\cii] \citep{papadopoulos2018new}.}
\mythirdcolor{It is important to mention that the distribution of [\ci] is influenced by
CRs, aligning it with the distribution of CO in the H$_2$ gas clouds \citep{bisbas2015effective,bisbas2017gmc}. This} contrasts to [\ci] being confined to a
narrow transition layer between the CO-rich inner $\rm H_{2}$ cloud areas and
the [\cii]-rich outer regions \citep{draine2010physics}, and  has
been supported by several studies, making [\ci] a robust tracer for the
molecular content of sources
\citep{papadopoulos2004greve,offner2014alternative,bisbas2021photodissociation,dunne2022dust}.

Ionized carbon, [\cii], can also serve as a molecular gas and SFR tracer \citep{olsen2017sigame,lagache2018cii,khusanova2021alpine,2021A&A...645A.133R,burgarella2022alma,glazer2024studying}. [\cii] has the advantage of being one of the
brightest lines that originate from star-forming regions
\citep{brauher2008compendium}. 
It is a result of the interaction between the
far-UV (FUV) photons and the interstellar medium (ISM) under typical ISM
conditions. \citet{velusamy2014origin}
and \citet{accurso2017radiative} showed that $\sim$ 60-85 $\%$ of the
total [\cii] emission arises from the molecular gas phase, which closely links
it with star formation regions. \mysecondcolor{As an SFR tracer}, [\cii] provides the advantage of being easily 
observed in a single frequency tuning,
compared to total infrared measurements, which require multi-wavelength
observations with different facilities. Despite the SFR -- [\cii] correlation, the so-called ``\cii $\,$ deficit'' is present. \mycolor{This effect refers to the weaker-than-expected
observed [\cii] emission from FIR in sources with enhanced star formation activity. Since [\cii], as one of the brightest FIR lines, is associated with the presence of FUV radiation, it would be expected to be as bright as the FIR continuum.} This ``\cii $\,$ deficit'' is a complex problem
that has been intensively investigated for more than a decade
\citep{gracia2011far,sargsyan2012c,casey2014dusty,lagache2018cii,bisbas2022origin,lahen2022panchromatic,10.1093/mnras/stad3792}.

This study aims to investigate the possible correlation between SFR and
$\rm L^{'}_{[CI]((1-0)}$ and $\rm L^{'}_{CO((1-0)}$ and 
\mycolor{the redshift dependency of these relationships}.
We use a large sample of sources \mycolor{(885 sources in total)} 
with redshifts spanning 0 < $\textit{z}$ < 6.5. We also discuss
the correlation of SFR with \mci, \mco, and \mycolor{lastly}, we dedicate a small
section discussing the SFR--$\rm L^{'}_{[CII]}$ correlation. Sect. \ref{sec:sample} presents and describes our sample compilation, providing details of the sources. In  Sect. \ref{sec:results} we present the results of our analysis. We further discuss the impact and reliability of the molecules used for the calculation of the SFR in Sect. \ref{sec:discussion}. Finally, we summarize our conclusions in Sect. \ref{sec:conclusions}. For this work, we have adopted a $\Lambda$CDM cosmology, with $\rm H_0$ = 67.8 km $\rm s^{-1} Mpc^{-1}$, $\rm \Omega_{M}$ = 0.308, and $\rm \Omega_{\Lambda}$ = 0.692 \citep{collaboration2014planck}, and a Salpeter initial mass function (IMF) \citep{salpeter1955luminosity}.

\section{Sample selection}\label{sec:sample}

\mysecondcolor{Our sample is a broad collection of sources from }
\citet{walter2011survey,alaghband2013using,sharon2016total,bothwell2017alma,valentino2018survey,valentino2020properties,harrington2021turbulent,montoya2023sensitive,dunne2021dust,dunne2022dust,berta2023,castillo2024comparative} (see Appendix \ref{appendix_total_sample} for details on the sample).
We also include a small sample of sources with [\cii] observations,
reported in \citet{cormier2015herschel,olsen2017sigame,glazer2024studying}. The
aforementioned works explore properties of the ISM, such as molecular mass,
luminosity, brightness temperature ratios, etc., including photodissociation region (PDR) properties,
such as the UV-radiation field
strengths, gas densities ($n$), gas temperatures, and possible scaling relations.
Statistics considering the lines used and the number of sources taken from each
work are presented in Appendix \ref{appendix_total_sample}. Finally, Fig.~\ref{fig:histograms2} displays the redshift distribution of our sample (only the CO and [\ci] samples (see below)).

To maintain a level of
homogeneity, all data were selected to have observations of both CO(1-0) and
[\ci](1-0) emission lines (except for the sources where we only used [\cii]
observations (i.e.,
\citet{cormier2015herschel,olsen2017sigame,glazer2024studying} samples)). \mysecondcolor{However}, 
South Pole Telescope (SPT) dusty star-forming galaxies (DSFGs) \citep{bothwell2017alma} and 
the \textit{z}-GAL survey galaxies \citep{berta2023}, \mysecondcolor{have only CO(2-1) data next to the }[\ci](1-0) 
observations. Also, the sample presented in \citet{valentino2020properties} \mysecondcolor{contains a small number of sources having only observations of CO(2-1) instead of CO(1-0}. Thus, for
a small number of galaxies used in this work (56 sources), the CO(2-1) transition was
utilized. \mysecondcolor{The CO(2-1) data were converted to
CO(1-0) intensities (see Sect. \ref{sec:luminosities} for more details)}.\mysecondcolor{We chose to also include the CO(2-1) transition since both CO(1-0) and CO(2-1) are emitted under similar gas temperature (T$\rm _{ex}$ $\sim$ 5.5, 16.6 K for CO(1-0) and CO(2-1), respectively) and density (n$\rm _{crit}$ = 2.2 $\times$ 10$^3$, 2.2 $\times$ 10$^4$ $\rm cm^{-3}$ for CO(1-0) and CO(2-1), respectively). Higher-J CO transitions (e.g., CO(3-2)) reported in the literature are not taken into account here, as they generally trace warmer and denser gas.}

Due to the nature of the sample, an overlapping listing effect is present across our 
literature samples (see details below). This overlapping listing effect has been taken into account in all calculations
presented here. Literature data taken include redshifts, magnification factors (in the case of lensed sources),
intensities ($\rm I_{[CI](1-0)}$, $\rm I_{CO(1-0)}$, $\rm I_{CO(2-1)}$, and
$\rm I_{[CII]}$), and total-infrared luminosities. Using those data quantities such as 
$\rm L^{'}_{[CI]((1-0)}$, $\rm L^{'}_{CO((1-0)}$, $\rm L^{'}_{[CII]}$, \mci, \mco, and SFRs were computed.

Considering the large sample utilized, a certain level
of inhomogeneity is expected (single-dish observations versus interferometric
observations, flux extraction, lensed sources, weighted mean of some measurements). 
The numbers presented in this
work and those mentioned in
\citet{walter2011survey,alaghband2013using,sharon2016total,bothwell2017alma,
valentino2018survey,valentino2020properties,harrington2021turbulent,
dunne2021dust,dunne2022dust,montoya2023sensitive,berta2023,castillo2024comparative}, could potentially 
have minor differences due to the selected cosmology or IMF choice.

Last, we present 
a rather small sample of sources with [\cii] observations
\citep{cormier2015herschel,olsen2017sigame,glazer2024studying}, along 
with the \citet{bothwell2017alma} sample that also includes [\cii] observational data.

\mycolor{Below}, a summary of the selected galaxy sample compilation is
presented. Despite the concise details provided below, we urge the reader to
refer to each work, as their authors provide information \mycolor{for values 
that will not be addressed in this work (e.g., brightness temperature ratios, conversion 
factor investigation, PDR modeling)}.

    From the combined sample of \citet{walter2011survey,alaghband2013using,sharon2016total}, eleven sources have been selected based on our criteria. Both CO(1-0) and [\ci](1-0) \mycolor{measurements were used}, by cross-matching the samples presented in \citet{walter2011survey} and \citet{sharon2016total}. Only one source (SMM J163650+4057) has an upper limit on the [\ci](1-0) line. This sample is comprised of typical sub-millimeter galaxies (SMGs) with redshifts span of \textit{z} $\sim$ 2.5-4. The observations were taken \mycolor{using the} IRAM 30m, IRAM Plateau de Bure, and VLA. Some of the sources have previously been presented in \citet{riechers2011co,riechers2011imaging,riechers2011extended,hodge2014kiloparsec}, using CO(1-0) or [\ci](1-0) observations separately. The majority of the sample (7/11) are gravitationally or strongly gravitationally lensed, with magnification factors of $\mu \, \sim$ 2.5-80. We note that previously reported upper limits of GN20 \citep{daddi2009two,casey2009search} have been replaced with the new IRAM NOEMA detection presented in \citet{cortzen2020deceptively}. In this work, we will refer to this sample as \citet{walter2011survey} for abbreviation, as they had the first publication reporting the majority of the sources.

    \citet{cormier2015herschel} sample comprises 43 low-metallicity star-forming (SF) galaxies of the guaranteed
    time key program Dwarf Galaxy Survey (DGS) conducted with the PACS instrument on \textit{Herschel}. Note that from the compact and extended samples presented in their work, only the compact sample was considered, as the latter sources were not fully mapped. The redshifts of this sample have a span of \textit{z} = 0.00034 - 0.04539.

    \citet{bothwell2017alma} sample comprises 13 strongly gravitationally lensed sources found in the 1.4 mm blank-field survey with the South Pole Telescope, with redshifts ranging between \textit{z} $\sim$ 3-4.7. \mycolor{These 13 sources were part of the Atacama Large Millimeter/submillimeter Array (ALMA) blind redshift search program \citep{weiss2013alma}, including observations of ground-state, low- and high-$J$ CO transitions.} In that program, 26 SPT DSFGs were observed, as part of the Cycle 0 `early compact array' setup, across the entirety of ALMA Band 3 (84-116 GHz). \citet{bothwell2017alma} reported the [\ci](1-0) transition, with only one source (SPT0345-47) as non-detection. In addition, two more sources (SPT0300-46 and SPT2103-60) have a tentative detection of $\sim$3$\sigma$ for the [\ci](1-0) transition. Sources from this sample do not have observations of the CO(1-0) transition; instead, the reported CO(2-1) transition was utilized for the purposes of this work \mycolor{(see Sect. \ref{sec:luminosities} for the method used to do the conversion)}. For this sample, we also used the reported [\cii] data.

    The sources taken from \citet{olsen2017sigame} were the 37 sources with \textit{z} $\geq$ 5 that had [\cii] observations. See \citet{olsen2017sigame} (and references therein) for further details in this sample, as the sources were previously reported in other works.

    \citet{valentino2018survey,valentino2020properties} sample comprises local-IR luminous galaxies, high-\textit{z} main-sequence galaxies, SMGs and QSOs (see references therein for further details of the individual samples this sample comprises). \mycolor{Because the \citet{valentino2018survey,valentino2020properties} sample includes sources from other samples utilized in our work (e.g., \citet{walter2011survey} sample), an adjustment had to be made to exclude double-counted sources.}
    Specifically, the local and high-\textit{z} sources reported in \citet{valentino2020properties} also include the \citet{valentino2018survey} sample in its entirety. \mycolor{The high-\textit{z} sources presented in \citet{walter2011survey,alaghband2013using,sharon2016total,bothwell2017alma} and four sources from the \citet{harrington2021turbulent} sample
    are also included in the high redshift sample of \citet{valentino2020properties}. 
    All the relevant sources from the aforementioned samples were excluded from the \citet{valentino2020properties} sample \mysecondcolor{used in the present work}.}
    This ensured that all sources used from \citet{valentino2020properties} were unique for comparison. 11/76 high-\textit{z} sources in \citet{valentino2020properties} had CO(2-1) observations \mysecondcolor{(see Sect. \ref{sec:luminosities} for details of the conversion).}
     Of the total of 217 sources, 76 have a [\ci](1-0) detection and 30 have a CO(1-0) detection, leaving the \mycolor{remainder} with either an upper limit or non-detection (see Appendix \ref{appendix_total_sample} for details). We also note that the sources taken from \citet{valentino2020properties} were cross matched between data presented in \citet{liu2015high} and \citet{kamenetzky2016vizier}. For sources with several estimates of the same low J-CO transition, the line flux was represented by a S/N-weighted average.

    \citet{harrington2016early,harrington2021turbulent} sample is part of the 24
    strongly lensed \textit{Planck} selected sources presented in \citet{harrington2021turbulent}. It was originally selected by the cross-match
    of \textit{Herschel} and \textit{Planck} 875 GHz detections
    \citep{harrington2016early}, and also 857, 545 and/or 353 GHz
    \textit{Planck} detections. In addition, these galaxies were chosen based on
    their high far-infrared luminosities ($\rm L_{IR} = (0.22-14.6) \times
    10^{12}\,\, \rm L_{\odot}$), which indicate high SFRs. 17/24 sources were selected that have
    \mycolor{both} [\ci](1-0) and CO(1-0) observations, with a redshift range of
    \textit{z} $\sim$ 1.1-3.5. 
    Previous results \citep{harrington2016early} suggest that despite the
    extreme $\rm L_{IR}$, most of it is due to high star formation activity and
    not dust-obscured AGN activity. Therefore, their characteristics suggest
    that the sources span over a region above main-sequence sources for their
    redshift values, classifying them as starburst objects.

    \citet{dunne2021dust,dunne2022dust} have an extensive sample \mysecondcolor{(312 sources in total)} that includes 
    high-\textit{z} SMGs, local star-forming galaxies, (U)LIRGs, normal \textit{z}=1 and 
    0.04 $<$ \textit{z} $<$ 0.35 galaxies. The sample contained lensed sources, sources with [\ci] data taken with the \textit{Herschel} Fourier Transform Spectrometer (FTS), and sources with only CO(2-1) observations. The \citet{dunne2022dust} sample includes also the 12 
    250-$\mu$m\ selected galaxies at \textit{z} = 0.35 reported in \citet{dunne2021dust}. 
    \citet{dunne2022dust} selected the sources from previous literature publications to have all three 
    main molecular gas tracers, namely CO(1-0), [\ci](1-0), and the submm dust continuum emission. 
    Several sources included in \citet{dunne2022dust} have also been reported in other samples used in this work. Namely, the sources presented in \citet{walter2011survey,bothwell2017alma,valentino2018survey,
    valentino2020properties,harrington2021turbulent} 
    were excluded from the \citet{dunne2022dust} sample \mysecondcolor{presented in this work}.

    \citet{montoya2023sensitive} sample comprises 40 local (U)LIRGs, 
    with a redshift range of \textit{z} $\sim$ 0.007-0.19. The sources had both new and 
    archival Atacama Pathfinder Experiment (APEX), and archival interferometric 
    ALMA/Morita Array (ACA) observations. \citet{montoya2023sensitive} selected 
    their sample based on previously reported \citep{veilleux2013fast,spoon2013diagnostics}
    \textit{Herschel} OH 119 \mum $\,$ observations, investigating molecular outflows. 
    16/40 sources and 22/40 have [\ci](1-0) and CO(1-0) observations, respectively.

     The \textit{z}-GAL sample (project IDs M18AB and D20AB; PIs: P. Cox, H. Dannerbauer, and T. Bakx) and the Pilot program (project IDs W17DM and S18CR; PI: A. Omont; \citealt{neri2020noema}), which together observed 137 \textit{Herschel}-selected sources with the IRAM Northern Extended Millimeter Array (NOEMA) and were presented in \citet{berta2023}, were also included in this work. The initial data release \citep{cox2023z} reported the survey details and the initial results (e.g., spectroscopic redshifts, right ascensions, declinations etc.). \citet{ismail2023z} continued by reporting the dust properties of the sample. 
    Several of the targets have been resolved into multiple objects. This accounts for the identification of \mysecondcolor{178} individual galaxies with a redshift span of 0.8 $<$ \textit{z} $<$ 6.5. \mysecondcolor{Five sources are excluded from this sample (HerBS-204, HerS-18 W, HerS-19 SE, HerS-19 W, and HerS-9), since information regarding their spectroscopic redshift was missing. This brings the total sample used in this work to 173 sources.} \citet {cox2023z} and \citet{berta2023} reported the [\ci](1-0) transition and additionally a selection of higher-J CO transitions from $\rm J_{up}$=2 to $\rm J_{up}$=8. Based on their reported data we only utilize the CO(2-1) transition (see Sect. \ref{sec:luminosities} for the method used to do the conversion), ignoring all of the higher-J transitions.

    \citet{castillo2024comparative} sample comprises CO(1-0) and [\ci](1-0) \textit{Karl G. Jansky} Very Large Array (JVLA) observations of 20 unlensed DSFGs at redshifts of \textit{z} = 2-5 and dust masses of $\rm M_d$ = 1-10 $\times$ $10^{9}$ $\rm M_{\odot}$. The sample is part of a total of 30 sources observed with JVLA \citep{castillo2023vla}. In the UKIDSS Ultra Deep Survey (UDS), Cosmological Evolution Survey (COSMOS), Chandra Deep Field North (CDFN), and Extended Groth Strip (EGS) fields from the S2CLS \citep{serjeant2017scuba} and S2COSMOS \citep{simpson2019east} surveys, a sample of sources found within 4 deg$^2$ of SCUBA-2 850 \mum\ imaging was used for the initial source selection. The brightest of those sources have additional ALMA and NOEMA observations \citep{stach2018alma,hill2018high,simpson2020alma,birkin2021alma,chen2022alma,liao2024alma}.

    \citet{glazer2024studying} sample presents new $\textit{z} \sim
    7$ [\cii] observations with ALMA, for three confirmed lensed Lya emitting
    galaxies. Of the three sources, only one had a 4-$\sigma$ detection of
    [\cii], while the remaining two reported with 3-$\sigma$ upper limits.
    Additionally, further observations of 6 < \textit{z} < 7 sources were
    included \citep{watson2015dusty,schaerer2015new,knudsen2016c,matthee2017alma, bradavc2017alma,
    carniani2018kiloparsec,bowler2018obscured,smit2018rotation,
    matthee2019resolved,hashimoto2019big,bakx2020alma,harikane2020large,wong2022alma,
    molyneux2022spectroscopic,fujimoto2024jwst,ferrara2022alma,schouws2023alma,heintz2023gas}.

\begin{figure}
\centering
\includegraphics[width = 3.5in]{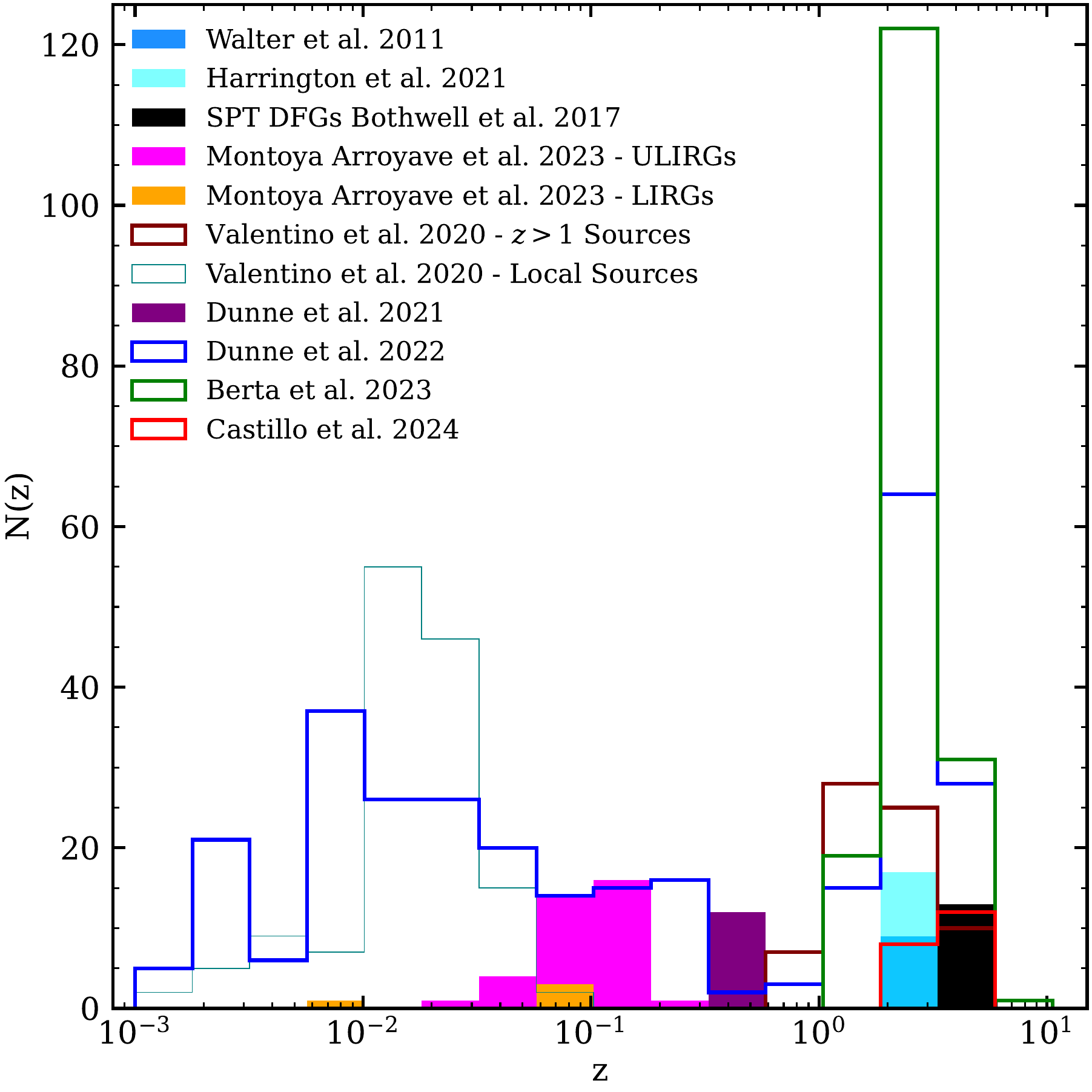}
\caption{Redshift distribution for our sample.}
\label{fig:histograms2}
\end{figure}

\section{Results and Analysis}\label{sec:results}

\subsection{Magnification corrections}

Strong gravitational lensing distorts the lensed source while increasing its
apparent brightness by a magnification factor ($\mu$), which depends both on the
mass of the intervening lens and the source/lens arrangement. As a large number
of sources reported in this work range from marginally gravitationally lensed to
strongly gravitationally lensed by a foreground galaxy, the effects of
gravitational lensing have to be quantified and accounted for. To account for
the magnification effect, when needed, we divided all computed numbers by a
magnification correction factor corresponding to each particular source. We must
note that most of the magnification factors used in this work have been computed
using 870-\mum\ data. Thus, we assume that the lensing models applied to these
sources also apply to the cold molecular gas traced by [\ci] and CO.

Due to the geometric nature of gravitational lensing, every individual location
within a lensed object is equally amplified across all wavelengths (see
\citealt{Bartelmann_2010} for a review). However, this does not imply a uniform
amplification across all wavelengths for the entire galaxy, since the extent of
the emission varies between different tracers.  Accurately determining the
magnification of dust sources presents a challenge. Submillimeter dust emission
occurs on larger scales, \mycolor{compared to molecular gas}, typically tens to 
hundreds of parsecs. Consequently, sources of this size tend to have lower 
overall magnifications, particularly
when the magnifications derived from optical observations are high (e.g., when
the optical source is near a caustic). When a source is large,
only a small portion of its area will be sufficiently close to a caustic to
experience significant amplification, while the rest will be farther away and
thus magnified to a lesser degree. This variation in magnification across
different parts of the source is referred to as `differential magnification' or
`chromatic lensing', and can introduce significant bias in the derived
properties of strongly lensed sources \citep{serjeant2012}. Additionally, the
finite extent of the background galaxy leads to variations in the magnification
applied to different regions within the galaxy. Consequently, the observed
spectral energy distribution (SED) \citep{blain1999differential}, as well as the
ratios of the spectral lines \citep{downes1995new,serjeant2012}, may be
distorted by this differential magnification if there are spatial variations in
the physical conditions within the source, as highlighted by
\citet{blandford1992cosmological}.

The phenomenon of differential lensing, which refers to the fluctuation of the
magnification factor across an extended source, can affect the properties of
strongly lensed sources. One consequence of differential lensing is that it
tends to bias the measurements of CO excitation toward more compact regions,
resulting in higher excitation levels, as noted by \citet{hezaveh2012size}. This
effect of differential lensing can distort certain apparent characteristics of
the source. For instance, if the impact of differential lensing is significant,
a lensed region with a temperature higher than the average will appear hotter
than its true temperature. However, low-$J$ transitions like CO(1-0) and
[\ci](1-0) are generally less influenced by the effects of differential lensing,
as they are well mixed with the molecular part of the ISM
\citep{ikeda2002distribution,papadopoulos2004c}. This indicates that the
underlying molecular gas and [\ci] or CO exhibit similar brightness profiles and
therefore their ratios remain unaffected by the effects of differential lensing.
Unfortunately, there is limited observational evidence that suggests the
presence of this impact on these transitions \citep{deane2013preferentially}.

It is worth mentioning that \citet{valentino2020properties} report two separate
values regarding the magnification factors for the high-\textit{z} sample, to
correct the continuum emission and its properties ($\mu_{\rm cont}$ and
$\mu_{\rm gas}$), but the mean difference ($\sim$1\%) is not significant. Here,
the gas magnification factors were adopted, for the corresponding sample as
suggested by \citet{valentino2020properties}, to study the properties of the ISM.
The magnification factors for all the other objects were taken from the
individual works that were first presented. We note that the sample reported in 
\citet{berta2023} does not have published magnification factors (Bakx et al. \textit{in prep}).
Since their sample possibly contains several gravitationally lensed sources,
a magnification correction is 
needed to correct for the lensing effects. For this reason, all sources taken from 
\citet{berta2023} are treated as upper limits (see Fig. \ref{fig:total_masses} , and Fig. \ref{fig:four_axes}). All quantities given below
($\rm L^{'}_{CO(1-0)}$, $\rm L^{'}_{[CI](1-0)}$, $\rm L^{'}_{[CII]}$, \mci,
\mco, and SFRs) have been de-magnified for lensing effects, wherever needed.
\footnote{To use a different magnification factor $\mu$, utilize the relation
$f_{\mu} = \mu/\mu_{\rm t}$, where $\mu_{\rm t}$ is the corresponding
magnification factor of each source adopted in this work.}

\subsection{Regression models} \label{sec:reg_models}

In this work, a power law fit was performed using two regression models. The
first model was implemented using the \emph{lmfit} package of \textit{Astropy}
\footnote{\citet{astropy:2013,astropy:2018,astropy:2022}}. It utilizes a
Levenberg-Marquardt algorithm \citep{gavin2019levenberg} provided in the
\emph{lmfit} package, that solves the least-squares problem accounting also for
the errors of the sources in both axes and fits a user-supplied model to the
supplied data. The second approach implements a Bayesian approach to linear
regression using the \emph{linmix}
\footnote{\url{https://github.com/jmeyers314/linmix}} package
\citep{kelly2007some}, which also accounts for the errors in both axes. Bayesian
reasoning is utilized, and a Markov chain is created with random samples from
the posterior distribution. The progress of the Monte Carlo Markov Chain (MCMC)
method toward the posterior distribution is assessed using the potential scale
reduction factor (RHAT), as introduced by \citet{gelman2004parameterization}.
Typically, a value of RHAT less than 1.1 indicates that an approximate
convergence has been achieved. The propagation of the error, for both models, on
the logarithmic axes have been taken into account as:

\begin{eqnarray} \label{eq:errprop}
    \rm \Delta F = \frac{1}{\ln{10}} \frac{\Delta x}{x},
\end{eqnarray}
where ln is the natural logarithm, $\rm \Delta x$ is the error of the corresponding
value where x denotes the actual value, and $\rm \Delta F$ is the final computed
error of the value.

\subsection{Line luminosities calculation}\label{sec:luminosities}

A small number of sources (see Sect. \ref{sec:sample}) lacked CO(1-0) observations. 
In those cases, the CO(2-1) transition was utilized for the
calculations. A brightness temperature ratio was used to convert the intensity of the CO(2-1) line to the intensity of CO(1-0), with
a value of $r_{21/10}\,=0.84\,\pm\,0.13$ \citep{bothwell2013survey,aravena2014co}.
Although CO(2-1) represents a higher excited state of CO (with \mycolor{a three times higher} critical density than CO(1-0), $n_{\rm crit}\approx 7
\times 10^3\,{\rm cm}^{-3}$ \citep{beaupuits2015}), it remains an excellent
tracer of the cold molecular gas, due to the low upper-level energy of 15 K. For this reason,
it can be converted without introducing significant errors to the corresponding CO(1-0)
intensities, \mycolor{assuming that the $r_{21/10}$ ratio remains approximately the same for different 
types of sources, and different environments. The latter assumption is in accordance to the recent numerical models of \citet{bisbas2021photodissociation} and within the errors of the 
value given in \citet{bothwell2013survey} and \citet{aravena2014co}}.  
\mycolor{In particular, \citet{bisbas2021photodissociation} report a ratio of $r_{21/10,\rm SF}/r_{21/10,\rm nSF}$ = 1.5, for the 
two simulated clouds; a star-forming and \mysecondcolor{a} non-star-forming \citep[see][]{wu2017gmc}.}

Following the work of \citet{solomon1997molecular,solomon2005molecular} we express the line luminosities in units of [K km s$^{-1}$ pc$^{2}$] as:

\begin{eqnarray} \label{eq:3.2}
    \rm L_{line}^{'} \,\,[K km s^{-1} pc^{2}] = 3.25 \times 10^{7} \,\nu_{rest}^{-2} \frac{D_L^2}{1+z} \int S_{v} du.
\end{eqnarray}

Here, $\nu_{\rm rest}$ is the rest-frame frequency of the particular line in GHz, $\rm D_L$ is the luminosity distance of the source, in Mpc, and $\int S_{v} du$ is the velocity ($u$) integrated flux of the observed line, in Jy~km~s$^{-1}$. Equation \eqref{eq:3.2} represents the line luminosities proportional to brightness temperature \citep{solomon2005molecular}.

\subsection{Total molecular mass and SFRs computations}\label{sec:m_mol_sfr}

Following the expression presented in \citet{papadopoulos2004greve,dunne2022dust}, the
[\ci](1-0) velocity integrated flux was used to calculate the total $\rm H_2$
mass (without the contribution of He):

\begin{eqnarray}\label{eq:3.4}
    \rm M(H_2)^{[CI]} \,\,[M_{\odot}] = \frac{0.0127}{\chi_{C} Q_{10}} \frac{D_L^2}{1+z} \int S_{[CI]} du
\end{eqnarray}
where $\chi_{\rm C}$ is the C/$\rm H_2$ abundance ratio, and $Q_{10}$ (see Appendix A in \citealt{papadopoulos2004c} for a detailed derivation of the equation excitation factor) is the fraction of carbon atoms in the first excited level, giving rise to [\ci](1-0) emission. The factor $\int S_{\rm [CI]} du$ represents the velocity integrated [\ci](1-0) flux.

The equation implies that all carbon emission stems from molecular gas, an
assumption that is justified by typical PDR models \citep{Ossenkopf2007} and
that allows us to compare the derived mass with the one obtained from CO. The
excitation parameter $Q_{10}$ of atomic carbon has the advantage of being quite
constant over a large range of temperatures and densities. A change in the
excitation in denser or hotter gas mainly shifts the level population from the
ground state to the upper state, leaving the intermediate one very stable.
\citet{papadopoulos2022subthermal} have shown that for typical ISM environments
($n_{\rm  H_2} = 300-10^{4}\,{\rm  cm}^{-3}$ and $T_{\rm  kin} = 25-80\,{\rm
K}$) a value of $Q_{10} = 0.48$ could be considered a typical value, as it
varies little ($\sim$16 per cent). 

The abundance of atomic carbon relative to molecular hydrogen, $\chi_{\rm C}$, 
depends on many physical parameters of the galaxies \citep{papadopoulos2004c}.
We followed a `standard' literature approach utilizing the value presented in
\citet{weiss2003gas} and \citet{papadopoulos2004greve}, adopting a C/$\rm H_2$
abundance ratio of $\chi_{\rm C} = 3 \times 10^{-5}$ (or $\rm \alpha_{CI} = 6.6 \,\,\msun (K \,km s^{-1} pc^2)^{-1}$) for our derivations, but
will discuss the parameter in more detail in Sect. \ref{highCI_ratio}. The use
of a consistent value for the abundance ratio of the corresponding molecule is a
crucial factor to calculate \mci and subsequently to make comparisons between
\mci and \mco.

The total $\rm H_2$ mass using the CO(1-0) velocity integrated flux was calculated using:

\begin{eqnarray}\label{eq:mco}
    \rm M(H_2)^{CO} \,\,[M_{\odot}] =  \alpha_{CO} L_{CO(1-0)}^{'}, 
\end{eqnarray}
and 

\begin{eqnarray}\label{eq:mci}
    \rm M(H_2)^{[CI]} \,\,[M_{\odot}] =  \alpha_{CI} L_{[CI](1-0)}^{'}, 
\end{eqnarray}

where $\rm \alpha_{CO}$ and $\rm \alpha_{CI}$ are the so-called CO-to-$\rm H_2$ and CI-to-$\rm H_2$ conversion factors in units
of $\rm \msun (K \,km s^{-1} pc^2)^{-1}$ \citep{solomon1997molecular}. Regarding
the conversion factor, \citet{bolatto2013co} have shown that it depends on
various physical effects such as temperature, metallicity, and cosmic rays. The
latter two environmental parameters can severely underestimate the value of the
conversion factor in the diffuse ISM as there is less dust shielding and lower
levels of CO self-shielding, while the dense ISM remains CO-rich and
bright. These effects lead to CO dissociation in the diffuse ISM, creating a
significant amount of $\rm H_2$ that is CO-dark
\citep[][\mysecondcolor{see also Sect. \ref{sec:ci_co_sfr_tracers} for more discussion on this}]{bisbas2015effective,bisbas2021photodissociation,
bisbas2024alpha,papadopoulos2018new}.

We \mysecondcolor{start here with} a CO-to-H$_2$ conversion factor of $\rm \alpha_{CO} =
0.8\,{M}_{\odot}\,({K}\,{km}\,{s}^{-1}\,{pc}^2)^{-1}$ \citep{downes1998rotating,bothwell2017alma,montoya2023sensitive} as most
of the studied sources are expected to have high (>100 $\rm M_{\odot} yr^{-1}$)
SFRs (see Sect. \ref{highCI_ratio} for a discussion on that). This value is slightly lower than the $\rm \alpha_{CO}\simeq1$ value commonly used for ULIRGs \citep{bolatto2013co}. This difference in the $\rm \alpha_{CO}$
factor introduces a $\sim$25\% increase difference on the individual \mco. It is clear that $\rm \alpha_{CO}$ has a large uncertainty from different studies. Recent studies even suggest a value of $\rm \alpha_{CO} =
4.0\,{M}_{\odot}\,({K}\,{km}\,{s}^{-1}\,{pc}^2)^{-1}$, based on a large cross-calibrated sample of sources \citep{dunne2022dust}. 
\mysecondcolor{Since this work does not aim to investigate conversion factors (as in \citet{dunne2022dust,Chiang2024,ramambason2024modeling}), the specific choice of $\rm \alpha_{CO}$ has little impact on our main premise.}

\begin{figure}
\centering
\includegraphics[width = 3.5in]{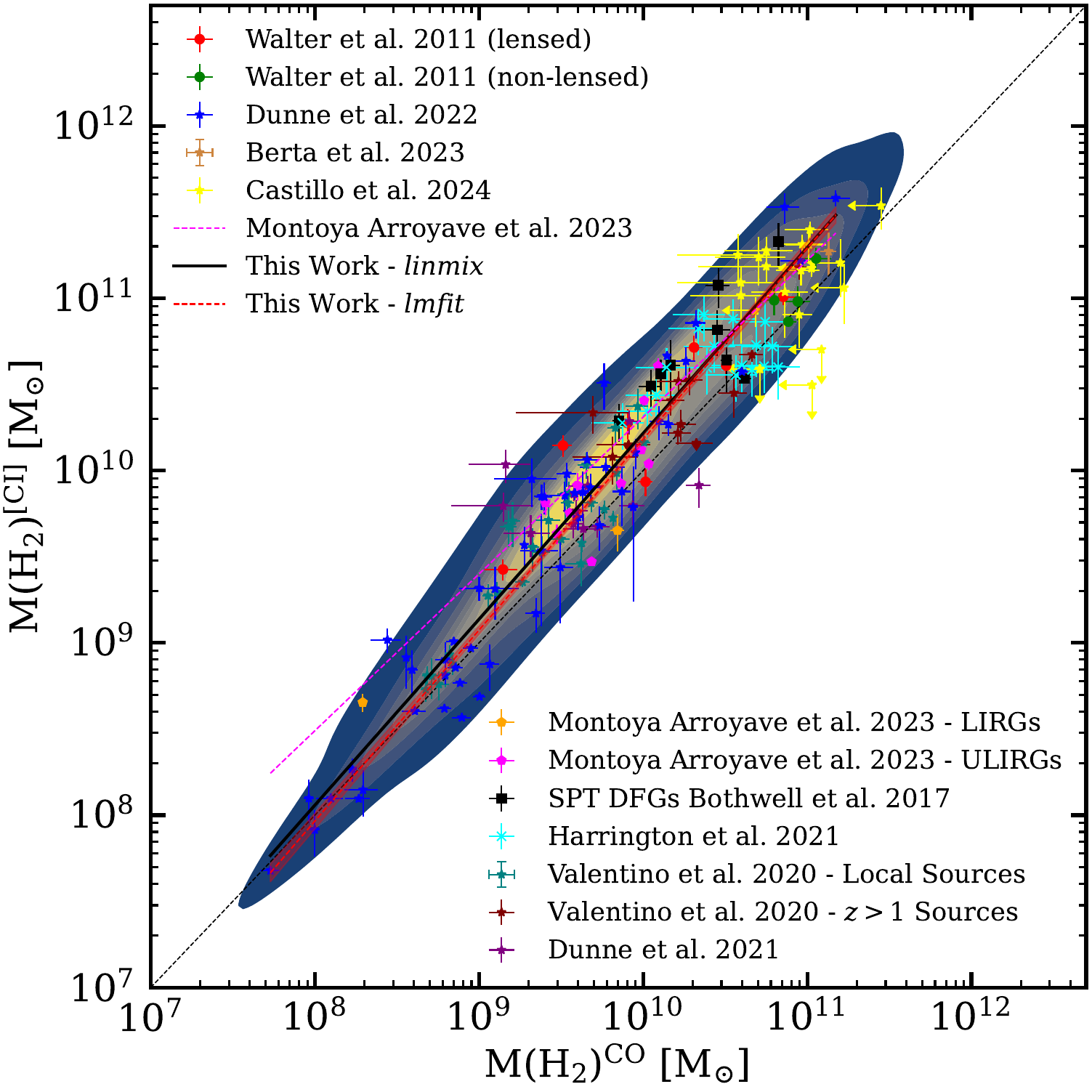}
\caption{\mci\ versus \mco\ plot. The best fit using the \emph{lmfit} package (red dashed line), and the \emph{linmix} package (thick black solid line) are presented in the figure. The 1-to-1 line is presented with a dashed black line. Finally, the relation derived in \citet{montoya2023sensitive} ($\rm \log M(H_2)^{[CI]} = (1.21 \pm 2.42) + (0.91 \pm 0.25)\,\rm \log M(H_2)^{CO}$) is included with a \mycolor{dotted} magenta line.}
\label{fig:total_masses}
\end{figure}

Figure \ref{fig:total_masses} shows a comparison of the calculated total
molecular masses \mysecondcolor{computed either from [\ci] or CO} using the two aforementioned conversion factors. We measure systematically
larger \mci\ compared to \mco\ for the same source. \mysecondcolor{This behavior
has also been observed even in rather small samples (e.g., \citealt{walter2011survey} and \citealt{bothwell2017alma}) with similar types of sources (SMGs and QSOs) as the ones presented in this work. This could suggest i) a higher C/$\rm H_2$ abundance ratio (assuming the same $\rm \alpha_{CO}$ factor) or ii) a higher $\rm \alpha_{CO}$ (assuming the same C/$\rm H_2$) to explain the mass difference. Our fits approach the 1-1 line if we increase the C/$\rm H_2$ to 5 $\times$ 10$^{-5}$ (i.e., by a factor of 1.7), or vise versa if we increase the $\rm \alpha_{CO}$ to 1.3 (i.e., by a factor of 0.6 ).

We also performed fits using the two methods presented in Sect. \ref{sec:reg_models}. All regression fits were performed using linear models in logarithmic space. Both fits are included in Fig. \ref{tab:slope_mass} with a red dashed line and a thick solid black line. The slope and intercept coefficients are presented in Table~\ref{tab:slope_mass}. They are almost identical within the errors. For comparison, the regression from \citet{montoya2023sensitive} is also included in Fig.~\ref{fig:total_masses}, showing only \mysecondcolor{a minor} deviation from our results.}

Star formation rates are computed using the \citet{kennicutt1998star} SFR--$\rm L_{IR}$ relation, assuming a Salpeter initial mass function (IMF):

\begin{eqnarray} \label{eq:3.6}
    \mathrm{SFR} \,\,[M_{\odot} yr^{-1}] = 1.71 \times 10^{-10} \rm L_{IR}, 
\end{eqnarray}

where $\rm L_{IR}$ is the total-IR luminosity integrated over 8-1000 \mum\ for
the corresponding source. We note that the \citet{harrington2021turbulent}
sources had no calculations of $\rm L_{IR}$ but rather far-infrared
luminosities, $\rm L_{FIR}$, integrated from 40-120 \mum\ spectra. To convert
the $\rm L_{FIR}$ into $\rm L_{IR}$ a similar method as presented in
\citet{stacey2021rocky} was followed. The reported $\rm L_{FIR}$
\citep{harrington2021turbulent} were multiplied by a factor of 1.91 ($\rm L_{IR}
= 1.91 \times L_{FIR}$), following \citet{dale2001infrared}, to account for
the mid-infrared spectral features. Equation \eqref{eq:3.6} was then used to
calculate the SFRs. We note that \citet{zhang2018stellar} found evidence that
starburst galaxies may contain a top-heavy IMF, different from the one assumed
in this work. This may overestimate the values of SFR derived here, as a
top-heavy IMF produces higher luminosities from the same mass of stars produced.
This issue has also been addressed in \citet{stacey2021rocky} for their sources.

The values of $\rm L_{IR}$ used in this work are primarily taken from
\citet{valentino2020properties}, except for the sources reported in
\citet{harrington2021turbulent,dunne2021dust,dunne2022dust,berta2023,montoya2023sensitive}, and \citet{castillo2024comparative}. In particular,
\citet{valentino2020properties} integrated (within 8-1000 \mum) 
the SED using data
available in the COSMOS spectroscopic coverage and a \citet{draine2007infrared}
model. They also corrected the $\rm L_{IR}$ of SMGs for the AGN contribution
(i.e., corrected for torus emission). Last, for QSOs, $\rm L_{IR}$ represents the
SFR and the AGN contribution. \mycolor{This could overestimate the 
$\rm L_{IR}$ for the QSO sources included in \citet{valentino2020properties} and subsequently our sample.
We finally note 
that the $\rm L_{IR}$ values for Cloverleaf and RX J0911+0551 \citep{walter2011survey} were taken from \citet{stacey2021rocky} and have been converted to $\rm L_{IR}$ via the method presented above. }

\mycolor{\citet{ismail2023z} measured the total infrared luminosity by integrating the modified black body (MBB) model for a range of 50-1000 $\mu$m. For this reason the \citet{berta2023} sources' L(50-1000 $\mu$m) were converted to $\rm L_{IR}$(8-1000 $\mu$m) using a L(50-1000 $\mu$m) over $\rm L_{IR}$(8-1000 $\mu$m) ratio of 0.7, derived using the SED template of \citet{berta2013panchromatic}.
We note here that the \citet{berta2023} sources are likely lensed, so the $\rm L_{IR}$(8-1000 $\mu$m) values could be overestimated due to lensing effects.}

\begin{figure*}
\centering
\includegraphics[width=3.5in]{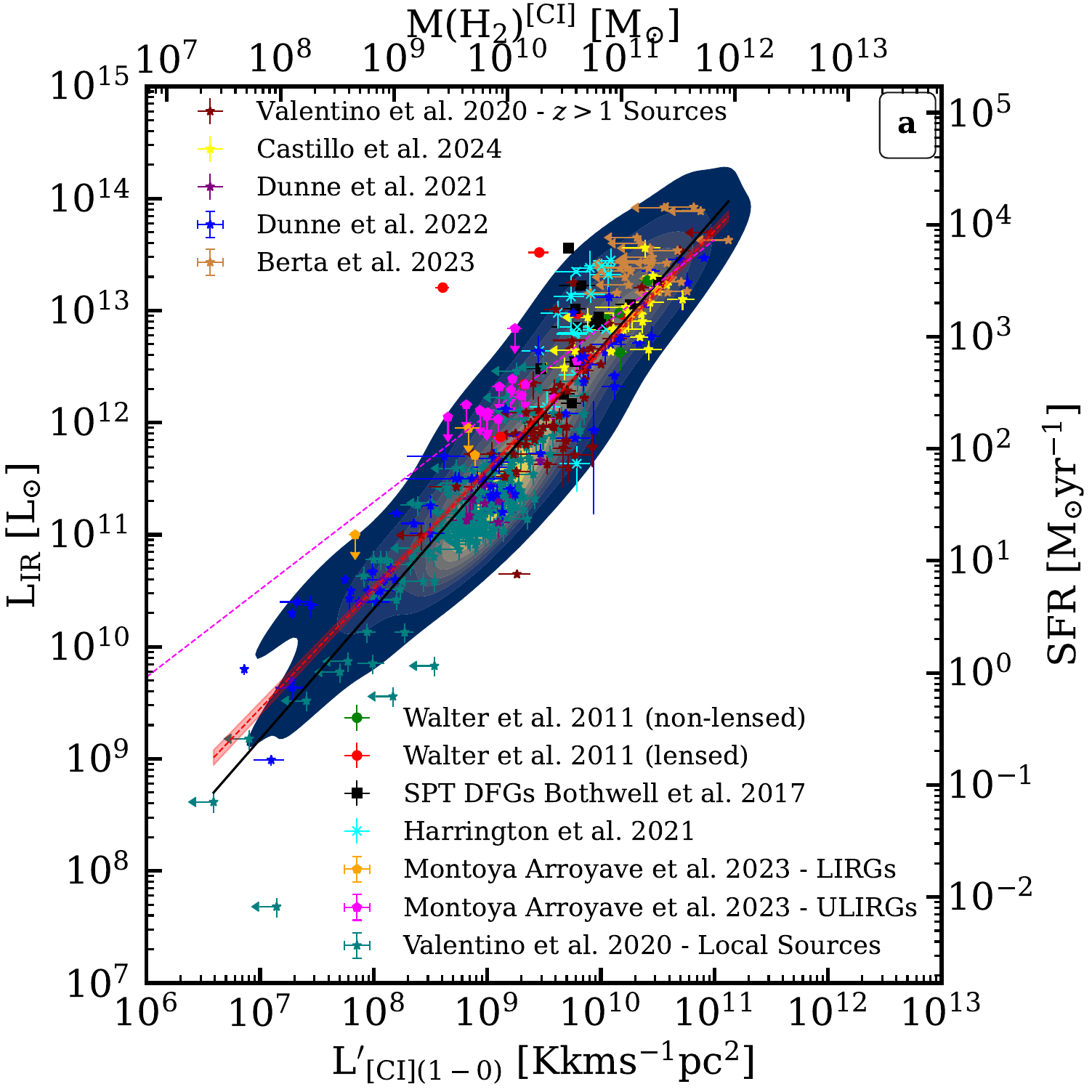}
\includegraphics[width=3.5in]{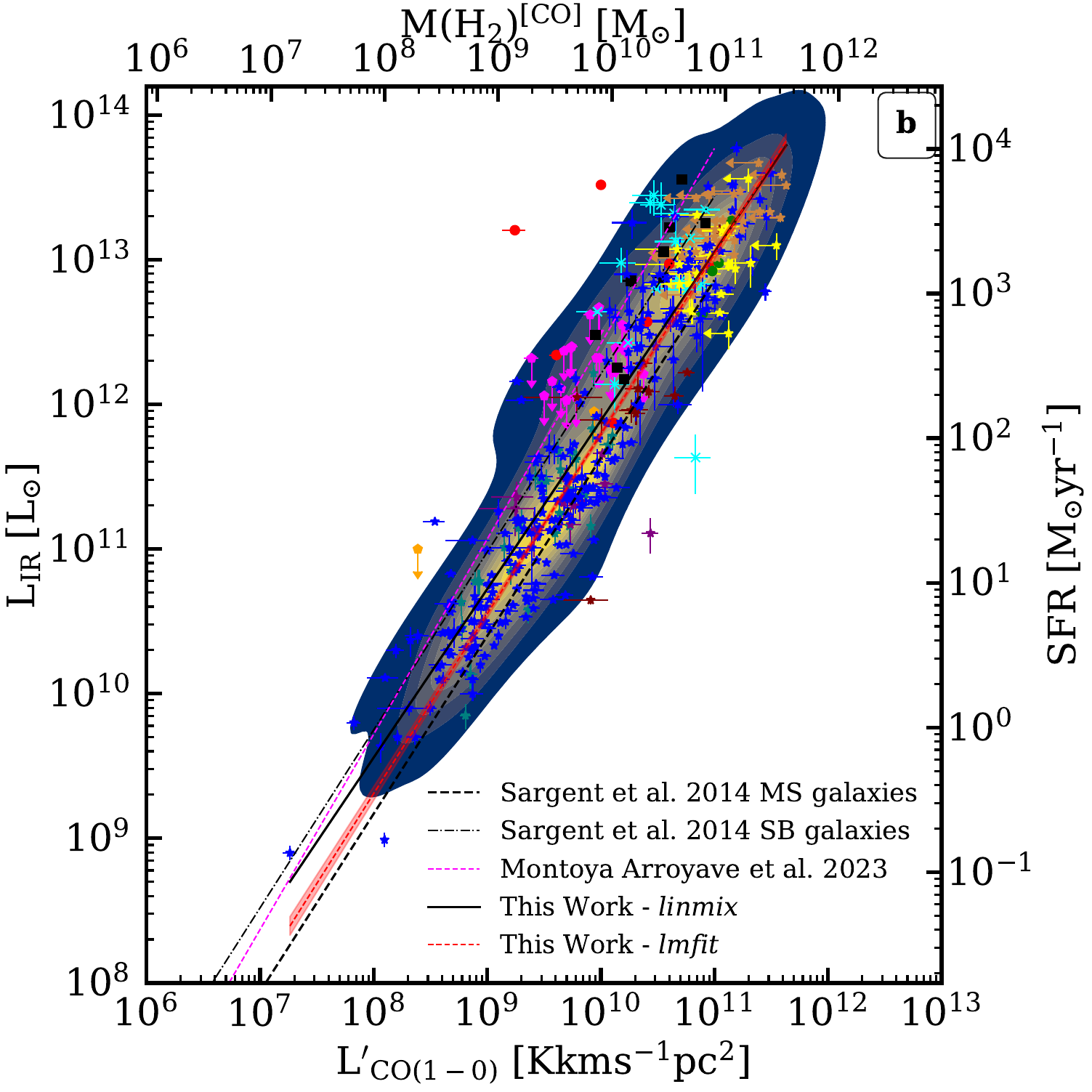}
\includegraphics[width=3.5in]{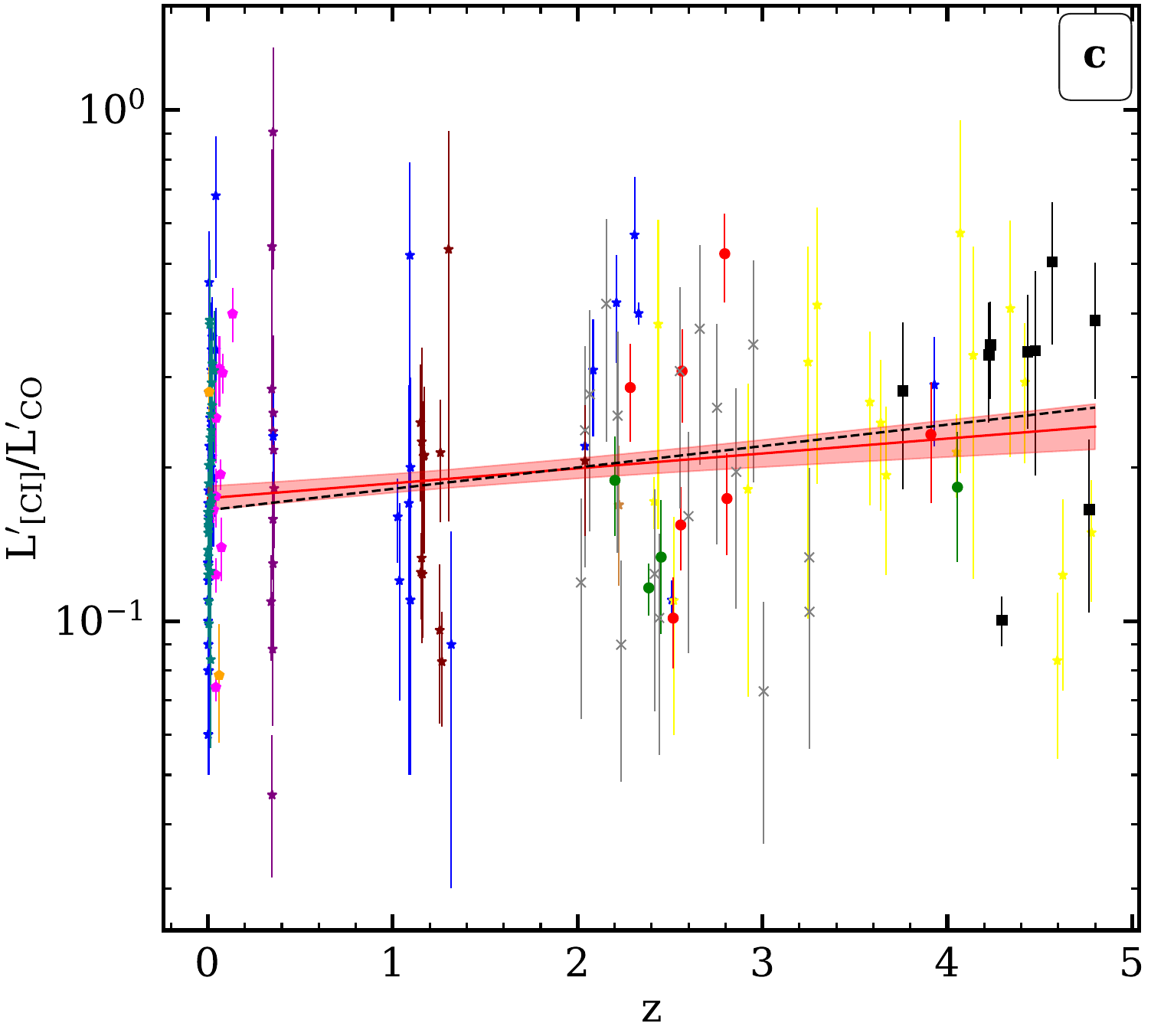}
\includegraphics[width=3.5in]{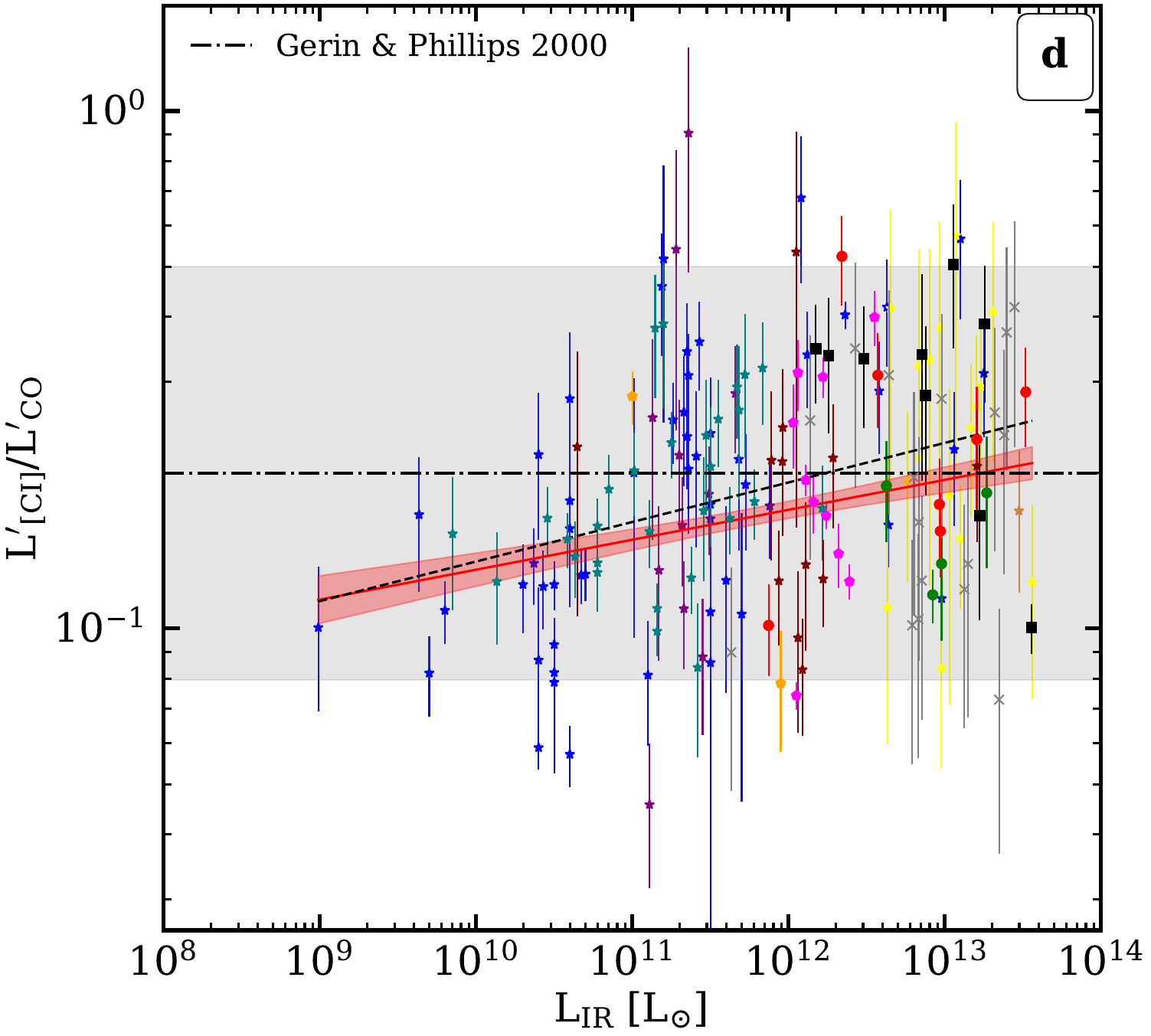}
\caption{$\rm L_{IR}$ versus $\rm L^{'}_{[CI](1-0)}$ (panel a), $\rm L^{'}_{CO(1-0)}$ (panel b) luminosities, $\rm L^{'}_{[CI](1-0)}$/$\rm L^{'}_{CO(1-0)}$ ratio against redshift (panel c), and $\rm L^{'}_{[CI](1-0)}$/$\rm L^{'}_{CO(1-0)}$ ratio against $\rm L_{IR}$ (panel d). The secondary y-axis represents the SFR and the secondary x-axis the total molecular masses computed using the corresponding tracer. We include the best fit using the \textit{Levenberg-Marquardt} algorithm of the \emph{lmfit} package (red dashed line), and the Bayesian approach to linear regression using the \emph{linmix} package (solid black line). In panel (b) we include the relations presented in \citet{sargent2014regularity} for main-sequence galaxies (thick black dashed line), and starburst galaxies (dashed-dotted line). In panels (a) and (b), the relations derived in \citet{montoya2023sensitive} for CO(1-0) and [\ci](1-0) line detections are presented (magenta dashed lines). Panels (c) and (d) include the \emph{lmfit} and \emph{linmix} method fits, with solid red and dashed black lines, respectively. \mysecondcolor{In panel (d), the black dash-dotted line presents the mean \mythirdcolor{$\rm L^{'}_{[CI](1-0)}$/$\rm L^{'}_{CO(1-0)}$} value and the scatter \mythirdcolor{(gray shaded area)} derived by \citet{gerin2000atomic}.}}
\label{fig:four_axes}
\end{figure*}

\begin{table}[ht]
\centering
\caption{Slope and intercept values for the \mci\ versus \mco\ relation.}
\label{tab:slope_mass}
\begin{tabular}{@{}ccc@{}}
\toprule
\toprule
Model  & \multicolumn{2}{c}{\mco} \\ \midrule
-      & slope ($\beta$)              & intercept ($\alpha$)      \\
lmfit  & 1.10 $\pm$ 0.03    & -0.90 $\pm$ 0.25        \\
linmix & 1.08 $\pm$ 0.02    & -0.57 $\pm$ 0.23      \\ \bottomrule
\end{tabular}%

\end{table}

\subsection{$\rm L^{'}_{[CI]}$ and $\rm L^{'}_{CO}$ correlation with total-IR luminosity and SFR}\label{sec:lum_correlation}

\mysecondcolor{Our sample comprises an extensive, large, but still heterogeneous sample of galaxies. The large number of sources should minimize observational biases that typically favour high $\rm L_{IR}$.}
While the independent observations can be biased (IR luminous sources,
gravitationally lensed sources, etc.) our work tries to minimize this bias (on a statistical level) 
by using a large sample \mysecondcolor{that contains} sources with values of
$\rm L_{IR}$ spanning $\sim$ 6 orders of magnitude (see Fig. \ref{fig:four_axes}). This
means that potential underlying galaxy scaling relations capture our sample in
its entirety, without biasing the derived linear regression slopes toward a
specific range of sources. \citet{montoya2023sensitive} discuss this \mycolor{bias
of small span in $\rm L_{IR}$} in their work, as their sample 
consisted only of local (U)LIRGs at similar redshifts (a similar discussion on
the scaling relations is also addressed in \citealt{cicone2017final}). \mycolor{We note that although our sample has a significant $\rm L_{IR}$ span, for higher 
redshifts it can be biased to very bright sources. This is because both tracers (CO and [\ci] lines) present observational 
challenges for sources with \textit{z} $>$ 4, due to high CR environments, turbulence, 
gas density, and metallicity \citep{bisbas2015effective,glover2016atomic,Bisbas2017,
papadopoulos2018new,clark2019tracing,bisbas2021photodissociation}.}

SFR and $\rm L_{IR}$ calibrations are produced via two regression fits based on observable patterns:
\begin{eqnarray} \label{eq:sfr}
     \rm logSFR = \alpha_{sfr} + \beta \,
     \rm logL^{'}_{line}, 
\end{eqnarray}
\begin{eqnarray} \label{eq:lir}
     \rm \log L_{IR} = \alpha_{IR} + \beta \, \rm logL^{'}_{line}, 
\end{eqnarray}
where SFR is the star formation rate of the source in units of $\rm M_{\odot} yr^{-1}$, $\rm L_{IR}$ is the total-IR luminosity of the source in units of $\rm L_{\odot}$, and $\rm L^{'}_{line}$ is the prime line luminosity of each tracer ([\ci](1-0), CO(1-0), or [\cii]) in units of K~km~s$^{-1}$~pc$^{2}$. Finally, $\beta$ and $\alpha$ are the slope and the intercept coefficients of the best fit, derived by one of the two used methods (see Sect. \ref{sec:reg_models}). Table \ref{tab:slopes} reports the derived values from the two models for [\ci](1-0), and CO(1-0).

\begin{table*}[ht]
\centering
\caption{Slope, intercept and dispersion values for the SFR (and $\rm L_{IR}$ in parenthesis) versus $\rm L^{'}_{[CI]((1-0)}$ and $\rm L^{'}_{CO((1-0)}$ relations.}
\label{tab:slopes}
\begin{tabular}{@{}cccccccc@{}}
\toprule
\toprule
Model & &\multicolumn{2}{c}{$\rm L^{'}_{[CI]((1-0)}$ [K~km~s$^{-1}$~pc$^{2}$]}     &    & \multicolumn{2}{c}{$\rm L^{'}_{CO((1-0)}$ [K~km~s$^{-1}$~pc$^{2}$]}           \\ \midrule
-      & slope ($\beta$) & $\alpha_{SFR}$ & $\alpha_{IR}$             & slope ($\beta$) & $\alpha_{SFR}$ & $\alpha_{IR}$      & dispersion           \\
lmfit  & 1.06 $\pm$ 0.02 & -7.76 $\pm$ 0.22 &\mysecondcolor{1.98} $\pm$ 0.21 & 1.24 $\pm$ 0.02 & -10.39 $\pm$ 0.24 & -0.63 $\pm$ 0.23 & 0.42 dex  \\
linmix & 1.16 $\pm$ 0.03 & -8.71 $\pm$ 0.26 & \mysecondcolor{1.16} $\pm$ \mysecondcolor{0.02}   & 1.17 $\pm$ 0.02 & -9.53 $\pm$ 0.24 & 0.24 $\pm$ 0.24  & 0.41 dex  \\ \bottomrule
\end{tabular}%
\vspace{0.1cm}

\small  Notes: Intercept values are given separately for the SFR and the IR luminosity (see Eq. \ref{eq:sfr} and \ref{eq:lir}). The dispersion values are the same regardless of the tracer. 
\end{table*}

Pearson correlation coefficients of 0.91 and 0.92 were derived for $\rm L_{IR}$
versus $\rm L^{'}_{[CI](1-0)}$ and $\rm L^{'}_{CO(1-0)}$ relations,
respectively. The \citet{montoya2023sensitive} relation ($\rm \log L^{'}_{[CI](1-0)} = (-6.42 \pm 4.91) + (1.27 \pm 0.40)\, \log L_{IR}$) was also
included in Panel (a) of Fig. \ref{fig:four_axes}, showing a shallower slope, compared to our values.
Regarding Panel (b) of Fig. \ref{fig:four_axes} the literature relations for
main-sequence ($\rm \log L^{'}_{CO(1-0)} = (0.54 \pm 0.02) + (0.81 \pm 0.03)\, \log
L_{IR}$) and starburst galaxies ($\rm \log L^{'}_{CO(1-0)} =
(0.08^{+0.15}_{-0.08}) + (0.81)\, \log L_{IR}$) presented in
\citet{sargent2014regularity}, and also the relation defined in
\citet{montoya2023sensitive} ($\rm \log L^{'}_{CO(1-0)} = (0.8 \pm 0.22) + (0.74
\pm 0.18)\, \log  L_{IR}$) were included. Even though not all
sources have enhanced star formation rates, even the starburst relation of
\citet{sargent2014regularity} agrees with a difference in slope only of $\sim 0.5
\%$  with the \textit{lmfit} model ($\sim 5 \%$ with \textit{linmix}).

Our derived relations for $\rm L_{IR}$ -- $\rm L^{'}_{CO(1-0)}$ lie between the
two relations presented in
\citet{sargent2014regularity} (MS and SB galaxies). As our sample contains both
main-sequence and starburst sources, it can be self-explanatory as to why this
is happening. The main-sequence and starburst relations of
\citet{sargent2014regularity} and the relation derived in
\citet{montoya2023sensitive} present a similar slope with the relations derived
here. Without taking into account the sources of the
\citet{montoya2023sensitive} sample, the main-sequence relation
\citep{sargent2014regularity} could be a good approximation for our sample
compilation, despite the small difference in slope with our derived relations.
On the contrary, the starburst relation is a better approximation for the
\citet{montoya2023sensitive} sample, as their similarity can be considered
negligible \mycolor{(the slope difference between \citet{montoya2023sensitive} relation 
and the SB relation from \citet{sargent2014regularity} is $\sim$ 9$\%$)}. 
This is also self-explanatory as all the \citet{montoya2023sensitive} sample 
comprises local (U)LIRGs.

\mycolor{Panels (c) and (d) of Fig. \ref{fig:four_axes} show the ratio of 
$\rm L^{'}_{[CI](1-0)}$/$\rm L^{'}_{\rm CO(1-0)}$ against redshift and $\rm L_{IR}$, respectively. 
Both panels display a broad range along the y-axis, with no clear indication of a distinct trend 
(panel (c): b = 0.029 $\pm$ 0.011, and 0.046 $\pm$ 0.017, for \textit{lmfit} 
and \textit{linmix}, respectively, panel (d): b = 0.057 $\pm$ 0.015, and 0.075 $\pm$ 0.017, for \textit{lmfit} and \textit{linmix}, respectively).
We derive a mean ratio of 
$\rm L^{'}_{[CI](1-0)}$/$\rm L^{'}_{\rm CO(1-0)}$ = 0.22 $\pm$ 0.07, \mysecondcolor{and Pearson correlation coefficients of 0.24 and 0.27 for panel (c) and (d), respectively. We also present the mean value \mythirdcolor{($\rm L^{'}_{[CI](1-0)}$/$\rm L^{'}_{\rm CO(1-0)}$ = 0.2 $\pm$ 0.2)} for a sample of local spirals, mergers, and low-metallicity galaxies, as described in \citet{gerin2000atomic}. Despite the large dispersion, our mean value is close to that of \citet{gerin2000atomic}}. [\ci](1-0) and CO(1-0) are strongly correlated independent of galaxy type and redshift, \mysecondcolor{ and a very weak dependence on $\rm L_{IR}$ only, given the large scatter}. Finally, although we expected some metallicity dependence, we found none, assuming that metallicity also evolves with redshift}.

Figure ~\ref{fig:four_axes} (panels (a) and (b)) shows that both $\rm L^{'}_{[CI](1-0)}$ 
and $\rm L^{'}_{ CO(1-0)}$ are tightly correlated with $\rm L_{IR}$
and SFR. We note that sources that had upper limits for SFR, $\rm L_{IR}$, $\rm L^{'}_{[CI](1-0)}$
and $\rm L^{'}_{CO(1-0)}$ were also taken into account for the
fitting. To account for these sources a conservative error of 20$\%$ of the reported value was
assumed (\mycolor{this will} be further discussed in Sect. \ref{sec:ci_co_sfr_tracers}).
\mycolor{This enabled us to include upper limits to the fitting
algorithms, thus utilizing the majority of sources from our sample.}

\subsection{SFR versus $\rm L^{'}_{[CII]}$ relation}\label{sec:cii_sfr}

\begin{figure}
\centering
\includegraphics[width = 3.5in]{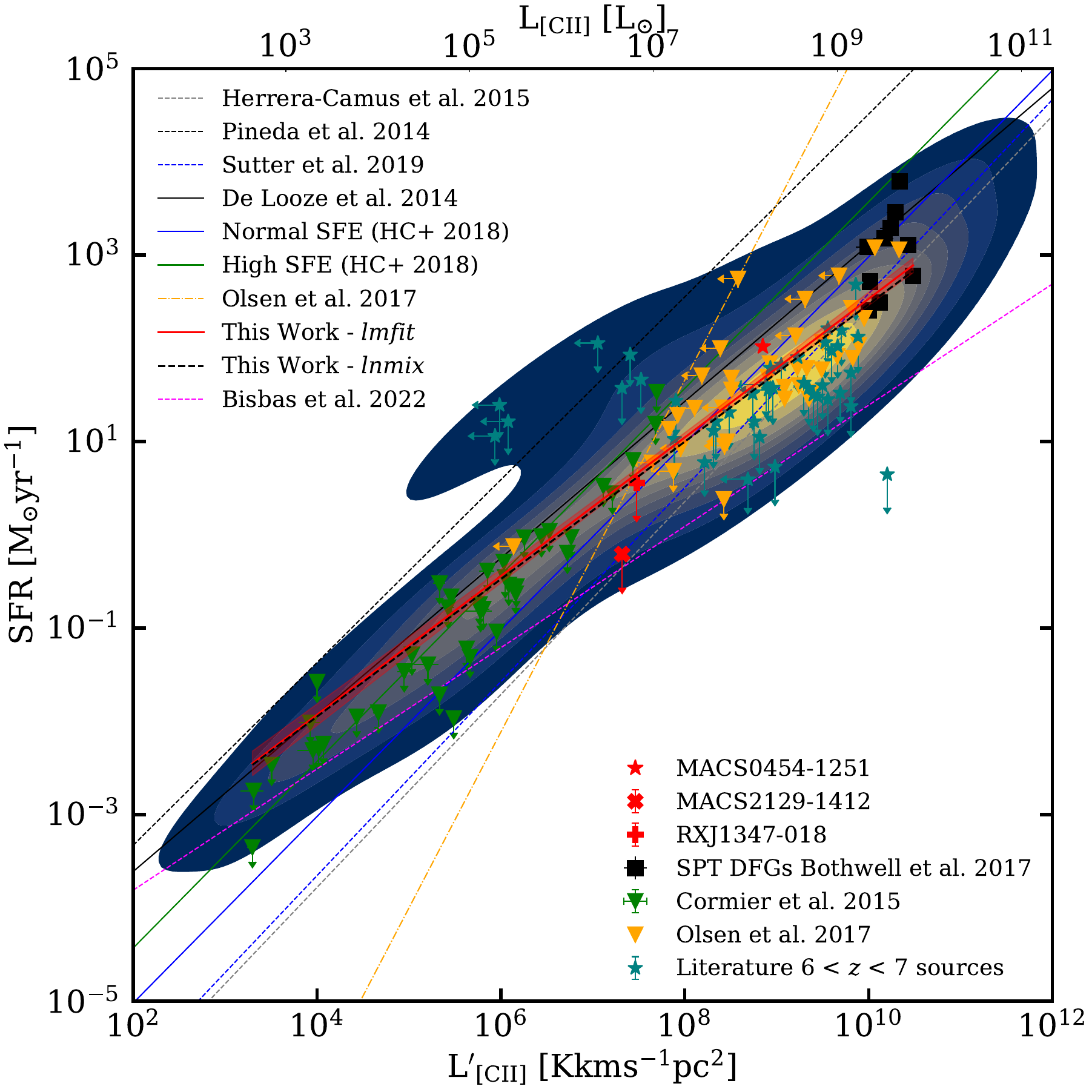}
\caption{SFR versus $\rm L^{'}_{[CII]}$ for the samples of \citet{cormier2015herschel} (green triangles), \citet{olsen2017sigame} (orange triangles), \citet{bothwell2017alma} (black squares) and \citet{glazer2024studying} and references therein (red symbols and teal stars). The red solid line represents the best fit for the \mycolor{listed sample} based on a Levenberg-Marquardt algorithm, and the thick black dashed line represents the Bayesian fitting. The gray, black, and blue dashed lines represent different best-fitting relations presented by \citet{herrera2015c,pineda2014aherschel}, and \citet{sutter2019using}, respectively. In addition, we plot with solid black, blue, and green lines the \citet{de2014applicability} Dwarf Galaxy Survey (DGS), the normal and high star formation efficiencies relations by \citet{herrera2018shining} (HC+ 2018), respectively. Also, the relation given in \citet{bisbas2022origin} is presented with a dashed magenta line. Finally, the relation that best fits the simulated data presented by \citet{olsen2017sigame} is also included in the figure with a dashed-doted yellow line.}
\label{fig:lcii_sfr}
\end{figure}

As mentioned in Sect. \ref{sec:into}, [\cii]\ also traces the SFR 
as it measures the UV irradiation from young massive stars independent
of the amount of molecular material. Although most of the sources used here
lacked [\cii]\ observations, the SPT sources \citep{bothwell2017alma} offered a
small sample to further extend previous fitting relations to larger values of SFR and
$\rm L^{'}_{[CII]}$. Additionally, literature observational data presented in
\citet{cormier2015herschel} and \citet{olsen2017sigame}, as well as recent ALMA
observations \citep{glazer2024studying} were included. We note that
\citet{olsen2017sigame} calculated their SFR using a Chabrier IMF, so their
reported numbers have been multiplied by a factor of 1.6, to account for the
Salpeter IMF used here. A best-fit relation was derived by using the two
aforementioned regression models (see Sect. \ref{sec:reg_models}), and
presented in Fig. \ref{fig:lcii_sfr} as a solid red and a black dashed line.

Figure \ref{fig:lcii_sfr} further includes literature best-fitting relations
\citep{de2014applicability,pineda2014aherschel,herrera2015c,olsen2017sigame,herrera2018shining,sutter2019using,bisbas2022origin}.
\citet{herrera2015c} reported in the \textit{Herschel} KINGFISH sample of 46
nearby galaxies, investigating the correlation of [\cii] surface brightness and
luminosity with the SFR of the corresponding sources. They followed an ordinary
least-squares (OLS) linear bisector method to best fit their data. This relation
is presented with a dashed gray line. \citet{pineda2014aherschel} investigated
the relation between [\cii] and SFR in the Galactic plane, by comparing their
results with the Large Magellanic Cloud (LMC), galaxies studies in
\citet{delooze2011}, and for individual PDRs. This relation is depicted with a
black dashed line. The blue dashed line corresponds to the relation derived by
\citet{sutter2019using}, who presented a sample of \textit{Herschel} KINGFISH
and Beyond the Peak programs data for 31 nearby sources that had both [\cii] and
[N{\sc ii}] 205 \mum\ spectral maps. The relation investigating the
applicability of [\cii] fine-structure line as a SFR tracer presented in
\citet{de2014applicability} is included in the best-fitting relations with a
solid black line. Both normal and high star formation efficiency (SFE) relations
presented in \citet{herrera2018shining} using \textit{Herschel}/PACS
observations of the main far-infrared (FIR) fine-structure lines are also
included, with solid blue and green lines, respectively. The relation that best
fits the simulated galactic data \citep{olsen2017sigame} produced with
\textit{SÍGAME} (SImulator of GAlaxy Millimeter/submillimeter Emission) is
included with the dashed yellow line. This relation is derived by fitting the
simulated multiphased ISM for 30 main-sequence galaxies at a redshift of
\textit{z} $\sim$ 6, with a rather small SFR ($\sim$ 3-23 $\rm
M_{\odot}$yr$^{-1}$). Those 30 simulated galaxies are not included in Fig.
\ref{fig:lcii_sfr}. Finally, the relation fitting the synthetic [\cii]
observations of smoothed particle hydrodynamics (SPH) simulations of a dwarf
galaxy merger presented in \citet{bisbas2022origin} is given with a dashed
magenta line. We note that the aforementioned literature fittings were performed
for SFR versus $\rm L_{[CII]}$ in solar luminosity units ($\rm L_{\odot}$). For this
reason, a complementary x-axis denoting $\rm L_{[CII]}$ was added in
Fig.\ref{fig:lcii_sfr}.

\begin{table}
\centering
\caption{Slope, intercept and dispersion values for the SFR versus $\rm L^{'}_{[CII]}$ relation.}
\label{tab:slope_cii}
\begin{tabular}{@{}cccc@{}}
\toprule
\toprule
Model  & \multicolumn{2}{c}{$\rm L^{'}_{[CII]}$} \\ \midrule
-      & slope ($\beta$)              & intercept ($\alpha$)  & dispersion    \\
lmfit  & 0.74 $\pm$ 0.02    & -4.93 $\pm$ 0.20    & 0.57 dex    \\
linmix & 0.73 $\pm$ 0.02    & -4.87 $\pm$ 0.22    & 0.57 dex  \\ \bottomrule
\end{tabular}%

\end{table}

The best-fitting relations from our models that correlate the sources presented in Fig. \ref{fig:lcii_sfr} are presented with a solid red and a thick black dashed line. A similar method was followed to the one previously used for SFR -- $\rm L^{'}_{[CI](1-0)}$ and SFR -- $\rm L^{'}_{CO(1-0)}$ (see Fig. \ref{fig:four_axes}), using the \emph{lmfit} and \emph{linmix} packages. The derived slope and intercept coefficients are presented in Table \ref{tab:slope_cii}.

From the literature best-fit relations presented in Fig. \ref{fig:lcii_sfr}, the one derived by \citet{de2014applicability} and the high SFE relation by \citet{herrera2018shining} have similar slopes with both our data and our derived relations (red solid and black dashed line). The \citet{de2014applicability} relation has a $\sim$13 \% difference in slope with the \textit{lmfit} relation and a $\sim$15 \% difference in slope with the \textit{linmix} relation. The slope of the high SFE relation by \citet{herrera2018shining} differs from the \textit{lmfit} and \textit{linmix} relations by $\sim$35 \% and $\sim$36 \%, respectively. The normal SFE relation of \citet{herrera2018shining} also presents merit, following a similar slope toward high $\rm L^{'}_{[CII]}$. This relation deviates to smaller SFRs toward the lower left part of Fig. \ref{fig:lcii_sfr}. Although a number of the aforementioned relations \citep{herrera2015c,sutter2019using} appear to give good results (less than $\sim$30\% difference with our models) toward the high SFR and high $\rm L^{'}_{[CII]}$ end of Fig. \ref{fig:lcii_sfr}, they present higher deviation at the low SFR and low $\rm L^{'}_{[CII]}$ end. The \citet{pineda2014aherschel} relation has the largest deviation for both high and low SFR, suggesting that it is not a good approximation for the sample used here (albeit it follows a prominent slope for the low SFR and low $\rm L^{'}_{[CII]}$ end of Fig. \ref{fig:lcii_sfr}, it fails to approximate the higher-end part).

The new observations presented in \citet{glazer2024studying} are also included in Fig.\ref{fig:lcii_sfr}. Of these three sources, only one (MACS0454-1251) had a 4$\sigma$ detection of [\cii], while the remaining two sources (RXJ1347-018 and MACS2129-1412) exhibit only upper limits. \citet{glazer2024studying} also discuss how low metallicity could shift galaxies below certain SFR--$\rm L^{'}_{[CII]}$ relations, as observed for the two upper limit galaxies presented in the corresponding work \mycolor{(and for a large majority of the upper limits utilized)}. This effect can also be observed for a large majority of \textit{z} > 6 sources that are included in Fig.\ref{fig:lcii_sfr} (teal stars, corresponding to galaxies with 6 < \textit{z} < 7). \mycolor{\citet{curti2024jades} addressed this effect and suggested that lower mass sources at redshifts \textit{z} $>$ 6 exhibit sub-solar metallicities, effectively destroying star-forming sites \citep{vallini2015c,ferrara2019physical}.} Our models also agree with the finding of \citet{glazer2024studying} regarding MACS2129-1412, concluding that this source is within 1$\sigma$ scatter from the \citet{de2014applicability} SFR--$\rm L^{'}_{[CII]}$ relation. Considering our modeling, both MACS0454-1251 and the upper limit for RXJ1347-018 ($\rm L_{[CII]} = 0.043 \times 10^{8}\,\, L_{\odot}$) are within $\sim$3$\sigma$ scatter from our two linear regression models. Following the one diverging source (MACS2129-1412), a large majority of the 6 < \textit{z} < 7 sources (teal stars) lie well below or above our models and additionally from the \citet{de2014applicability} relation, suggesting that the [\cii] deficit could play a more significant role for these sources \citep{vallini2015c,bisbas2022origin}. However, the majority of the sources are upper limits on either SFR and/or $\rm L^{'}_{[CII]}$. Despite our derived relations agree with previously reported ones, their use of mostly upper limits as an input makes them unreliable for this sample.

\section{Discussion}\label{sec:discussion}

\subsection{Total molecular masses and abundance ratios} \label{highCI_ratio} 
A value for the C/H$_2$ ($\chi_{\rm C}$) abundance ratio of $\chi_{\rm C}$=3$\times10^{-5}$  \citep{papadopoulos2004greve,bothwell2017alma,montoya2023sensitive} was used to compute the total molecular gas mass via the [\ci] line. This value of the abundance ratio is the average of the minimums recorded in the Orion A and B clouds ($\sim 10^{-5}$, \citealt{ikeda2002distribution}) and in the starburst environment of M82 nucleus (5$\times \, 10^{-5}$, \citealt{white1994co}). \mycolor{The recent study of \citet{jiao2021carbon} examined} a sample of six nearby galaxies and suggested that the commonly adopted value for $\chi_{\rm C}$ is well within the statistical variance of the sources examined in their corresponding work. Although \citet{jiao2021carbon} (and references therein) investigated only a small number of nearby sources, their derived value (C/H$_2=2.3\times10^{-5}$) agrees well with the value commonly adopted in the literature (i.e., 3 $\times$ 10$^{-5}$). It will be useful, however, to explore how different values of this ratio affect the results.

A commonly used technique to benchmark two different tracers in estimating the total molecular gas mass is to keep one of the two fixed while changes are performed in the other one. Following this, by keeping $\rm \alpha_{CO}$ fixed at $0.8\,{\rm M}_{\odot}\,({\rm K}\,{\rm km}\,{\rm s}^{-1}\,{\rm pc}^2)^{-1}$ we vary the C/H$_2$ ratio (or $\rm \alpha_{CI}$, see Eq. \ref{eq:mci}) and re-calculate the H$_2$ gas masses traced by [\ci]. 
\mysecondcolor{As a large number of sources in our sample have relatively enhanced $\rm L_{CO(1-0)}^{'}$, our choice of $\rm \alpha_{CO}$ = 0.8 is a fair assumption. The opposite approach using a different $\rm \alpha_{CO}$ will be discussed below.} We find a good match of the masses if we increase the C/H$_2$ ratio by a factor of 1.6, so that C/H$_2=5\times10^{-5}$. This approach was also considered in \citet{bothwell2017alma} finding a better agreement between the CO- and [\ci]-traced H$_2$ gas masses by using an elevated C/H$_2 \sim7\times10^{-5}$. \citet{castillo2024comparative} followed a similar approach, deriving an abundance ratio of $\chi_{\rm C}$ = 5.6 $\pm$ 2.0 $\times$ 10$^{-5}$ and $\chi_{\rm C}$ = 5.1 $\pm$ 1.8 $\times$ 10$^{-5}$ when they separate their sample into two different redshift bins (\textit{z} < 3.5, and \textit{z} > 3.5). Evidently, an elevated C/H$_2$ ratio of $5\times10^{-5}$ may be preferred for these types of galaxies, especially those with enhanced SFRs. 

There are several reasons to expect elevated abundances of C in the ISM of these galaxies. High SFR environments would suggest high CR and a more turbulent environment \citep{Papadopoulos2010}. CRs can penetrate deep into a cloud \citep{Strong2007,Padovani2009,Grenier2015}, making it C-rich via reactions \mycolor{caused by} the presence of high abundances of He$^+$ and H$_3^+$, ions that are produced from the CR interaction \citep{Bialy2015,bisbas2015effective,Bisbas2017,Gaches2022}. In addition, the astrochemical models of \citet{Bisbas2023} showed that the molecular mass content in column-density distributions \mycolor{similar to} a star-forming ISM, is C-rich across a large span of CR ionization rates and metallicities, and even C-dominant in environments of a low FUV intensity. It is noted here, however, that the more recent PDR and radiative transfer models representing an $\alpha$-enhanced low-metallicity star-forming cloud of \citet{bisbas2024alpha}, show that low C/O ratios can severely suppress the [\ci] emission while boosting the low-$J$ CO one, leaving the H$_2$ abundance in the cloud unaffected. This suggests that a potential $\alpha$-enhanced environment may have a significant effect in some of the targets presented here. It can affect the [\ci] emission and subsequently the [\ci] computed total molecular mass, without affecting the total SFR of the corresponding sources \citep[see][as an example of a {[}\ci{]}-dark galaxy]{Michiyama2021}.

\mysecondcolor{Instead of modifying the atomic carbon abundance, there are also arguments for a higher $\rm \alpha_{CO}$ value than used here. Using a large heterogeneous sample (407 sources) of dusty star-forming galaxies, \citet{dunne2022dust} found an $\rm \alpha_{CO}$ = 4.0 $\pm$ 0.1 $\rm {M}_{\odot}\,({K}\,{km}\,{s}^{-1}\,{pc}^2)^{-1}$ (including helium) and a C/$\rm H_2$ abundance ratio of 1.6 $\times$ $10^{-5}$. Since their sample is selected to have sources with FIR/sub-mm/mm wavelength it can be safely assumed that they have high- or comparable metallicities with the Milky Way, highlighting that 
metallicity can severely affect both the $\rm \alpha_{CO}$ conversion factor and the
C/$\rm H_2$ abundance ratio values (\citealt{jiao2021carbon,Bisbas2023, bisbas2024alpha,bisbas2025metallicity}). While these authors support those near-universal values 
for both $\rm \alpha_{CO}$ and $\rm \chi_{C}$ factors, many high-resolution studies of gas
kinematics in SMGs cannot easily support this high value for $\rm \alpha_{CO}$ based on 
their derived stellar and dynamical masses \citep{magdis2011goods,riechers2020coldz,birkin2021alma,castillo2022kiloparsec,amvrosiadis2025kinematics}. Using $\rm \chi_{C} = 5 \times 10^{-5}$ (or $\rm \alpha_{CI} = 3.94 \,\,\rm {M}_{\odot}\,({K}\,{km}\,{s}^{-1}\,{pc}^2)^{-1}$), and assuming $\rm \alpha_{CO} = 0.8$, yields consistent H$_2$ masses derived from both CO and [\ci] across the sample. The resulting values imply that the majority of sources in our sample have solar or super-solar metallicities, as noted by \citet{heintz2020direct}.}

The $\rm \alpha_{CO}$ conversion factor has been found to increase at \mycolor{low metallicity} \citep{leroy2011co,genzel2012metallicity,Amorin2016,shi2016carbon} depending also on the density distribution \citep{schruba2017physical,bisbas2021photodissociation}. In line \mycolor{with} these findings are the recent works of \citet{Chiang2024} and \citet{ramambason2024modeling}. In particular, \citet{Chiang2024} found $\rm \alpha_{CO}$ = 4.2 $\rm {M}_{\odot}\,({K}\,{km}\,{s}^{-1}\,{pc}^2)^{-1}$ using dust emission as a tracer of the gas surface density. \citet{ramambason2024modeling} performed representative modeling of the ISM using the Bayesian code MULTIGRIS in combination with low-metallicity dwarf galaxies observations and found that CO may not be a reliable H$_2$ gas tracer in low metallicity environments. The [\cii] fine-structure line can be used as an alternative tracer in such cases \citep{Zanella18}, and including the ISM environments of high CR ionization rates, as the $\alpha_{\rm CII}$ conversion factor has been found to not strongly dependent on metallicity \citep{cormier2015herschel,madden2020tracing,bisbas2021photodissociation,hunter2024interstellar}. In this regard, observations in the [\cii] emission line \mycolor{are} recently of popular interest, since it can be detected by ALMA in the high-$z$ Universe.

\mysecondcolor{Adopting an $\rm \alpha_{CO}$ = 4.0 $\rm {M}_{\odot}\,({K}\,{km}\,{s}^{-1}\,{pc}^2)^{-1}$, as suggested by \citet{dunne2022dust}, we find that the resulting total molecular masses require a C/$\rm H_2$ abundance ratio of 1.2 $\times$ 10$^{-5}$ to remain consistent. We calculate an $\rm \alpha_{CI}$ $\sim$ 17 $\rm {M}_{\odot}\,({K}\,{km}\,{s}^{-1}\,{pc}^2)^{-1}$, a value similar to the one derived in \citet{dunne2022dust}. Despite this change in $\rm \alpha_{CO}$ and C/$\rm H_2$ abundance ratio, the resulting \mythirdcolor{[\ci]-based and CO-based total molecular masses} fit derived in Fig. \ref{fig:total_masses} still exhibits a super-linear behavior with a slope of $\sim$ 1.10, consistent with our previous analysis.}

\subsection{Comparison of $\rm L^{'}_{CO(1-0)}$-\LIR\ and $\rm L^{'}_{[CI](1-0)}$-\LIR\ with the literature.} \label{sec:ci_co_sfr_tracers}

Taking into account the assumptions and corrections that have been implemented
to compute the total SFR and also the $\rm L^{'}_{CO(1-0)}$ and $\rm L^{'}_{[CI](1-0)}$ of the sources, we derived a close-to-linear
correlation with SFR (see Sect. \ref{sec:lum_correlation}). We remind the
reader that the SFR was computed for the majority of the sample using 8-1000
\mum\ integrated total-infrared luminosity observations that were acquired using
detailed SED integration \citep{valentino2020properties,dunne2022dust,berta2023}. The sources presented
in \citet{harrington2021turbulent} had 40-120 \mum\ integrated FIR observations,
therefore an extrapolation similar to the one presented in
\citet{stacey2021rocky} was used to convert from $\rm L_{FIR}$ to $\rm L_{IR}$
(see Sect. \ref{sec:m_mol_sfr}). We also \mycolor{stress} that possible IMF
normalizations were taken into account for our computations. 
\mycolor{We provide a reminder of these assumptions} as they are 
crucial in making meaningful and consistent comparisons of a 
large dataset such as the one used in this work.

Using a sample of 885 sources, we have shown that both [\ci](1-0) and CO(1-0)
luminosities correlate with the total SFR, over approximately 6 and 5 orders of
magnitude, respectively (see Table \ref{tab:slopes}). With a span in 
redshift covering 0 < \textit{z} < 6.5,
empirical relations with a slope of \textit{b} = 1.06 $\pm$ 0.02 using the
\textit{lmfit} model (\textit{b} = 1.16 $\pm$ 0.03 using the \textit{linmix}
model) was derived for the SFR--$\rm L^{'}_{\rm [CI](1-0)}$ relation. Similar
coefficients were derived for the SFR--$\rm L^{'}_{CO(1-0)}$ transition. A slope of
\textit{b} = 1.24 $\pm$ 0.02 was derived using the \textit{lmfit} model
(\textit{b} = 1.17 $\pm$ 0.02 using the \textit{linmix} model). As the final
relations partially depend on upper limits that were taken into account for the
fitting, a 20$\%$ error based on the corresponding original value was assumed;
10\% following the \citet{galametz2013calibration} calibrations and an
additional conservative 10\% due to e.g., flux calibration errors, supernovae
remnants that could potentially contribute to $\rm L_{IR}$, and mergers
\citep{hopkins2010mergers}. \mycolor{Although similar correlations have been discussed in previous works 
\citep{valentino2018survey,valentino2020properties,montoya2023sensitive},
they were biased mostly toward the number and type of the adopted sources. 
Considering the potential bias of those previous work,
their derived relations give similar trends, albeit with different slope values.}

Based on a sample of 36 galaxies,  \citet{montoya2023sensitive} provide a
correlation between SFR and $\rm L^{'}_{[CI](1-0)}$ or $\rm L^{'}_{CO(1-0)}$
when no LIRGs contribution is taken into account, for a SFR range of
$\sim 17- 1183\,{\rm M}_{\odot}\,{\rm yr}^{-1}$.
We note that those correlations correspond to SFRs computed
using a different method from the one utilized here. In particular, the authors calculated/taken 
literature values of SFR that use
the expression presented in \citet{sturm2011massive} and in
\citet{spoon2013diagnostics} (namely ${\rm SFR} = (1- \rm \alpha _{AGN}) \times
10^{-10} \rm L_{IR}$, where $\rm \alpha_{AGN}$ is the AGN contribution factor) accounting 
also for the AGN contribution. By using the richer sample
of galaxies presented here, we find that this correlation (albeit with slightly
different slope) can be used for a wider range of SFRs. This, in turn,
implies that a meaningful correlation between SFR and $\rm L^{'}_{[CI](1-0)}$ or
$\rm L^{'}_{CO(1-0)}$ can be derived with a wide span of values in redshift and $\rm L_{IR}$.

\mycolor{Although} the SFR--$\rm L^{'}_{CO(1-0)}$ relation has been extensively studied in
the literature
\citep{gao2004star,bayet200912co,juneau2009enhanced,greve2014star}, there is a
lack of similar works examining the SFR versus $\rm L^{'}_{[CI](1-0)}$
correlation. In addition, although extensive samples have been reported using
CO(1-0) or [\ci](1-0) observations to investigate the molecular content and/or
the SFR
\citep{leroy2005molecular,lisenfeld2011amiga,cicone2017final,valentino2018survey,valentino2020properties,harrington2021turbulent,dunne2021dust,dunne2022dust,berta2023,castillo2023vla},
to our knowledge no comprehensive study combining both these tracers to such \mycolor{wide} redshift span has been previously made.
Our study compares total molecular mass content and the SFR
using both tracers from the same sources, thus minimize possible \mycolor{redshift} biases.
However, we have to address that the majority of the sources were selected using IR- and FIR-bright \textit{Herschel} data, resulting in a large number of (U)LIRGs, starbursts, or low-metallicity dwarf galaxies. This can introduce a bias toward only the IR- and FIR-bright sources across our redshift range, especially for redshifts \textit{z} $>$ 4, \mysecondcolor{as also explained in Sect. \ref{sec:lum_correlation}}.

\citet{bothwell2017alma}, \citet{valentino2020properties} and
\citet{montoya2023sensitive} suggested that [\ci] can give reliable results of 
the total molecular mass content of galaxies, and can be thus used as an
alternative molecular gas tracer for local and extragalactic sources
\citep{papadopoulos2004c,lo2014tracing,zhang2014physical}. 
Only recently and for a limited number of sources, \citet{bothwell2017alma} and
\citet{montoya2023sensitive} compared \mci\ and \mco\ and found a larger
molecular content when using [\ci](1-0). Figure \ref{fig:total_masses} illustrates the
larger sample selection used in this work and, as can be seen, agrees with their results.
This supports the hypothesis that [\ci](1-0) \mycolor{could} trace larger quantities of cold
molecular gas in local and high-\textit{z} galaxies, compared to CO(1-0).

The discrepancy in \mci\ and \mco\ can also be explained by the `CO-dark' gas \citep{vDishoeck90,vDishoeck92}. 
The term `CO-dark' refers to molecular regions that due to the lower self-shielding ability of CO  
compared to \Htwo, can remain \Htwo-rich but be CO-poor \citep{Lada88}. 
This `CO-dark' gas can be manifested as either due to the lack of CO molecules, 
existing in translucent or diffuse clouds in molecular regions due to the weaker
CO emission \citep{Seifried20}, or due to the low sensitivity limit of telescopes to detect the 
CO(1-0) rotational transition \citep{langer2014}. A weaker CO emission can be compensated by raising the value of the $\rm \alpha_{CO}$ conversion factor. However, since there is no information on the column density of CO, N(CO), we are unable to correct for the 'CO-dark' discrepancy using the $\rm \alpha_{CO}$.

Furthermore, the cosmic microwave background radiation (CMB) 
can have a severe impact on the computed total 
molecular mass using the CO(1-0) \mysecondcolor{transition}
\citep{da2013effect,zhang2016gone} for \textit{z}>4. 
This stems from the fact that the CMB increases in temperature and therefore it becomes
brighter at higher-\textit{z}. Low-$J$ CO lines, because of their low frequency, would be decreased after its radiation transfer with the CMB,
showing much less contrast than their intrinsic line emission. This is especially severe for molecular gas with
kinetic temperature below 20 K, while for warm gas with $\rm T_{kin} \, \sim$ 50 K, this effect would decrease the CO line
luminosity within 50\% \citep{zhang2016gone}. [\ci] lines are much less affected by this CMB effect and therefore
would be more robust for the high redshift Universe. 
The CMB effect increases both $\rm \alpha_{CO}$ and $\rm \alpha_{CI}$, \mysecondcolor{subsequently decreasing} the corresponding tracer emission. 
This reduction in emission was explored in \citet{castillo2024comparative}, 
which proposed a non-negligible 
decrease in both [\ci](1-0) and CO(1-0) emission. However, this decrease appears to have a counteracting 
effect on the measured luminosity ratio, as reported by \citet{castillo2024comparative}. Literature values considering high-\textit{z} SMGs and ULIRGs, report 
log ($\rm L^{'}_{[CI](1-0)}$/$\rm L^{'}_{\rm CO(1-0)}$) = -0.71 $\pm$ 0.12, 
while \citet{valentino2018survey} report a value 
of log ($\rm L^{'}_{[CI](1-0)}$/$\rm L^{'}_{\rm CO(1-0)}$) = -0.69 $\pm$ 0.16. Both literature 
ratios and our calculated one (see Sect.\ref{sec:lum_correlation}) are in close agreement. Furthermore, \citet{jarugula2021molecular} and \citet{harrington2021turbulent} have measured high kinetic temperatures ($\rm T_{dust}$ $\sim$ 45 K, $\rm T_{kin}$/$\rm T_{dust}$ $\approx$ 2.5) for a selection of strongly lensed sources with \textit{z} $>$ 4. This could explain the relatively minor effect of CMB in our high-\textit{z} sample (slopes $<$ 0.1, see Sect. \ref{sec:lum_correlation}, and Fig. \ref{fig:four_axes}).

\subsection{[\cii]\ as a SFR tracer}
 
Numerous studies explore the ability of [\cii] in tracing the SFR of galaxies \citep{de2014applicability,pineda2014aherschel,cormier2015herschel,olsen2017sigame,herrera2018shining,sutter2019using,glazer2024studying}. Here, we further extend the aforementioned studies by adding the SPT sources \citep{bothwell2017alma}, and the more recent observations reported in \citet{glazer2024studying}, the majority of the latter being only upper limits. 
The [\cii] sample utilized in this work is a rather small collection of low SFR galaxies \citep{cormier2015herschel}, intermediate to high SFR sources \citep{olsen2017sigame,glazer2024studying} and high SFR sources \citep{bothwell2017alma}. Our derived SFR--$\rm L^{'}_{[CII]}$ correlation (slopes of \textit{b} = 0.74 $\pm$ 0.02 and \textit{b} = 0.73 $\pm$ 0.02, for the \textit{lmfit} and \textit{linmix} model, respectively) agrees well (see Sect. \ref{sec:cii_sfr}) with the previously reported ones \citep{de2014applicability,herrera2018shining}, suggesting the tight correlation of [\cii] with SFR \citep{lagache2018cii}. Figure~\ref{fig:lcii_sfr} shows this correlation.

\citet{de2014applicability} follow a similar approach to investigate the applicability of [\cii] line as a SFR tracer. Our slope agrees to within $\sim6\%$ with their result, although our dispersion is somehow larger ($\sim0.64$ dex) than the \citet{de2014applicability} one ($\sim0.42$ dex) due to the smaller sample of galaxies we use.

\subsection{Dependency on the Initial Mass Function for \textit{z}>8}

With the advent of JWST, further observations of high-redshift sources
(\textit{z}>8) will become available in the future. It is, thus, important to
consider changes in the IMF as a function of redshift to correct for the stellar
mass and SFRs \citep[e.g.,][]{steinhardt2023templates,hennebelle2024physical}. These new templates
take into account the vastly different high-\textit{z} Universe, bringing in
agreement previously derived results \citep{boylan2023stress} with the
$\Lambda$CDM cosmology. As \cite{sneppen2022implications} have shown in the
COSMOS project, while a change in the IMF has negligible effects for
\textit{z}<4 galaxies, their physical parameters depend on the IMF shape for
\textit{z}>8 and suffer if a Galactic IMF is used. 

\cite{steinhardt2023templates} have also discussed this problem by analyzing
recent JWST data of \textit{z}>8 galaxies. They conclude that as the
highest-\textit{z} galaxies are found as dropouts with the NIRCam photometry
lacking narrow band detections to constrain the Balmer break, the only method to
constrain the best-fit redshift would be based on the equivalent width of that
partial detection. However, a change in the UV slope will change the equivalent
width and subsequently make the best-fit for the redshift sensitive to IMF
changes. Considering that, \mycolor{should further sources with 
\textit{z} $>$ 8 be added in the aforementioned sample, it is recommended to 
pay significant attention to potentially different IMF slopes. Therefore a switch from 
the Salpeter IMF used in this work to a more appropriate one for very high
\textit{z} sources (e.g., Chabrier) might be needed.}

\section{Conclusions}\label{sec:conclusions}
The SFR--$\rm L^{'}_{[CI](1-0)}$,  SFR--$\rm L^{'}_{CO(1-0)}$ and SFR--$\rm L^{'}_{[CII]}$ relations have been explored in this work using a data sample of 885 galaxies spanning a redshift of $0<z<6.5$, found in the literature (see Appendix \ref{appendix_total_sample} for more details of the sample). 
Using two simple regression models (\emph{lmfit} and \emph{linmix}) following two different statistical methods (\textit{Levenberg-Marquardt} and Bayesian reasoning) we derived a super-linear correlation for SFR--$\rm L^{'}_{[CI](1-0)}$, and SFR--$\rm L^{'}_{CO(1-0)}$, with the latter having a \mycolor{steeper} slope. The two models have negligible differences in the sample span without hinting at any over- or under-fitting. Considering the small dispersion numbers, for both approaches, it is clear that the linear regression models are not biased toward any redshift data, even though the sample has a larger collection of low- to mid-\textit{z} sources. The main findings of this work are summarized as follows: 

\begin{itemize}
    \item  
    A systematic difference in the molecular mass content is found if [\ci](1-0) is used as an alternative tracer to CO(1-0). In particular, we derive higher amounts of cold molecular gas using [\ci] than CO. 
    This difference in mass could potentially hint toward a `CO-dark' H$_2$-rich gas in our overall sample. It could also hint toward underestimating \mysecondcolor{either} the $\rm \alpha_{CI}$, \mysecondcolor{or the $\rm \alpha_{CO}$ factor} assumed here. By keeping the $\rm \alpha_{CO}$ factor constant and using a smaller value of $\rm \alpha_{CI}$ = 3.9 $\rm {M}_{\odot}\,({K}\,{km}\,{s}^{-1}\,{pc}^2)^{-1}$ we can bring the two masses in agreement. This strengthens the hypothesis that [\ci] can be used as a reliable tracer of the bulk of $\rm H_2$ gas in galaxies. \mysecondcolor{By adopting a value of $\rm \alpha_{CO}$ = 4.0 $\rm {M}_{\odot}\,({K}\,{km}\,{s}^{-1}\,{pc}^2)^{-1}$ we would deduce a conversion factor of $\rm \alpha_{CI}$ $\sim$ 17 $\rm {M}_{\odot}\,({K}\,{km}\,{s}^{-1}\,{pc}^2)^{-1}$. Both of these factors are directly related. Independent of this fundamental uncertainty, we observe a slightly super-linear trend in the derived fit.}

    \item The \emph{lmfit} model that utilizes a \textit{Levenberg-Marquardt} algorithm to solve the least squares problem gives a super-linear slope with $\beta = 1.06 \pm 0.02$ ($\beta = 1.24 \pm 0.02$) for the SFR--$\rm L^{'}_{[CI](1-0)}$ (SFR--$\rm L^{'}_{CO(1-0)}$) relation. Similar results are derived using the \emph{linmix} Bayesian reasoning regression model. A super-linear correlation with a slope of $\beta = 1.16 \pm 0.03 $ ($\beta = 1.17 \pm 0.02$) was derived for SFR--$\rm L^{'}_{[CI](1-0)}$ (SFR-$\rm L^{'}_{CO(1-0)}$) relation. Finally, using a small sample to explore the SFR--$\rm L^{'}_{[CII]}$ relation we derived a correlation with slope $\beta = 0.74 \pm 0.02$ ($\beta = 0.73 \pm 0.02$) by using the \emph{lmfit} (\emph{linmix}) model.

    \item The Pearson coefficients of 0.91 for the SFR--$\rm L^{'}_{[CI](1-0)}$, 0.92 for the SFR--$\rm L^{'}_{CO(1-0)}$, and 0.92 for the SFR--$\rm L^{'}_{[CII]}$ relations suggest a tight correlation for all three tracers. Despite the large coefficient for the $\rm L^{'}_{[CII]}$ the majority of the data are upper limits, suggesting caution for this high value. Comparing [\ci] and CO coefficients, both give equally good results, with the latter having an insignificantly higher correlation value.  

    \mycolor{\item The derived slopes suggest that SFR scales superlinearly with $\rm  L^{'}_{[CI](1-0)}$ ($\rm L^{'}_{CO(1-0)}$). The use of CO and CI as SFR tracers overcomes biases that are known to exist when using [\cii] as a tracer, making CO and [\ci] more robust.}

\end{itemize}

Overall, our analysis supports that [\ci] and CO can be used as a reliable SFR tracers for a \mycolor{wide} span in redshift. The \mycolor{small dispersion}, in combination with the high Pearson coefficients, results in relations that can be used across redshift to compute the SFR using [\ci] or CO (see Table \ref{tab:slopes}). New observations of low- and high-\textit{z} sources would significantly impact the derived relations, as it would help better understand how the presented relations behave in low- and high-SFR environments. Additional sources with low and high values of $\rm L^{'}_{[CI](1-0)}$ or $\rm L^{'}_{CO(1-0)}$ would help strengthen our findings by pushing the derived relations to the [\ci] or CO luminosity limit.


\begin{acknowledgements}
    We thank the anonymous referee for their useful
    suggestions and prompt comments which improved the quality of this manuscript. 
      Part of this work was supported by the German
      \emph{Deut\-sche For\-schungs\-ge\-mein\-schaft, DFG\/} project
      number Ts~17/2--1. T.T. gratefully acknowledges the Collaborative Research Center 1601 (SFB 1601 subproject A6) funded by the Deutsche Forschungsgemeinschaft (DFG, German Research Foundation) – 500700252. T.T also thanks Dominik A. Riechers for all the helpful conversations and insights. We acknowledge the support of the National Natural Science Foundation of China (NSFC) under grants No. 12041305, 12173016. We acknowledge the Program for Innovative Talents, Entrepreneur in Jiangsu.  We acknowledge the science research grants from the China Manned Space Project with NOs.CMS-CSST-2021-A08 and CMS-CSST-2021-A07. We acknowledge support from the Leading Innovation and Entrepreneurship Team of Zhejiang Province of China (Grant No. 2023R01008).
\end{acknowledgements}



\bibliographystyle{aa}
\bibliography{bibliography}

\begin{appendix}

\section{SFR against $\mathrm L{'}_{line}$ with redshift dependence} 

\begin{figure}[ht]
\centering
\includegraphics[width = 3.5in]{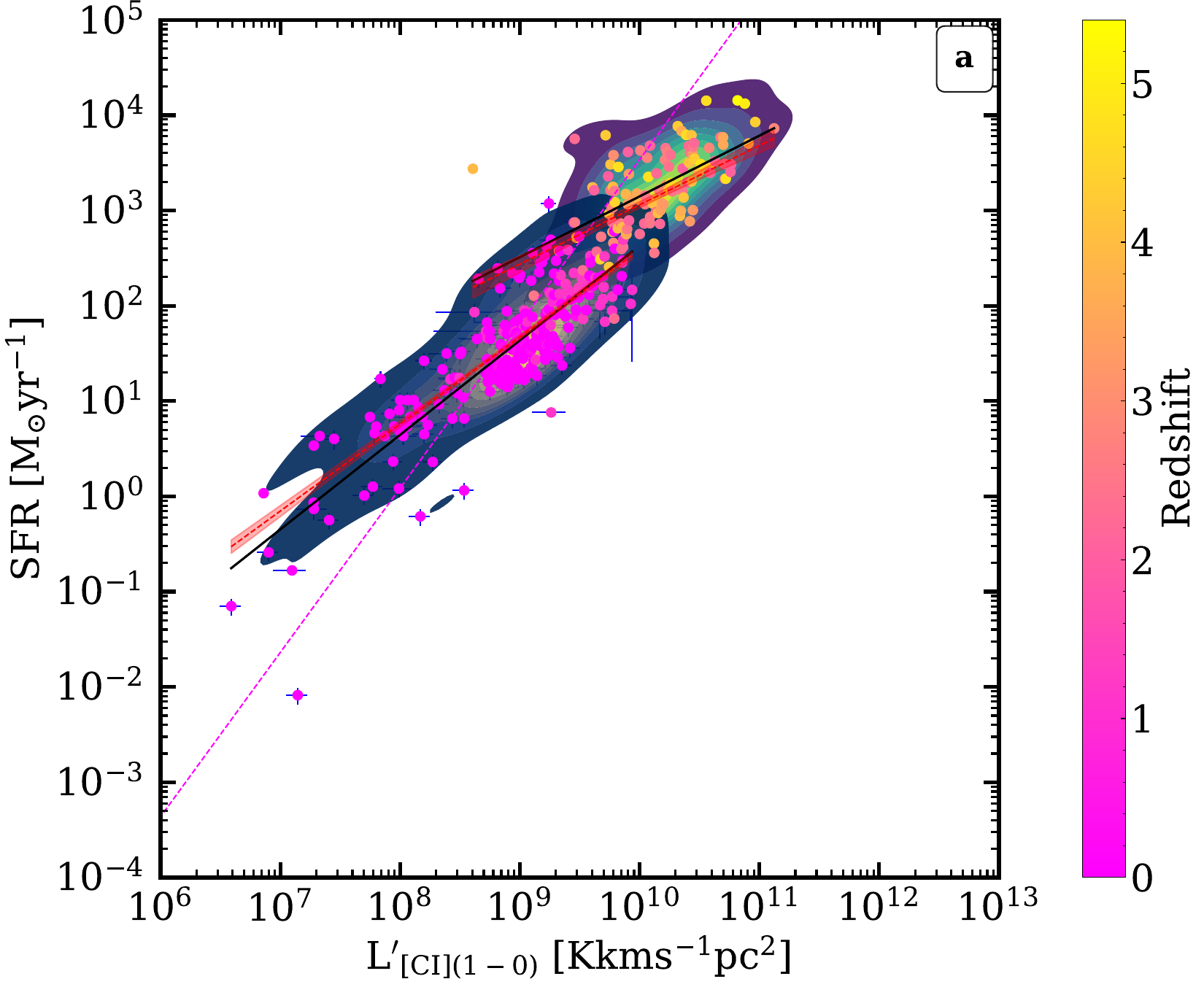}
\includegraphics[width = 3.5in]{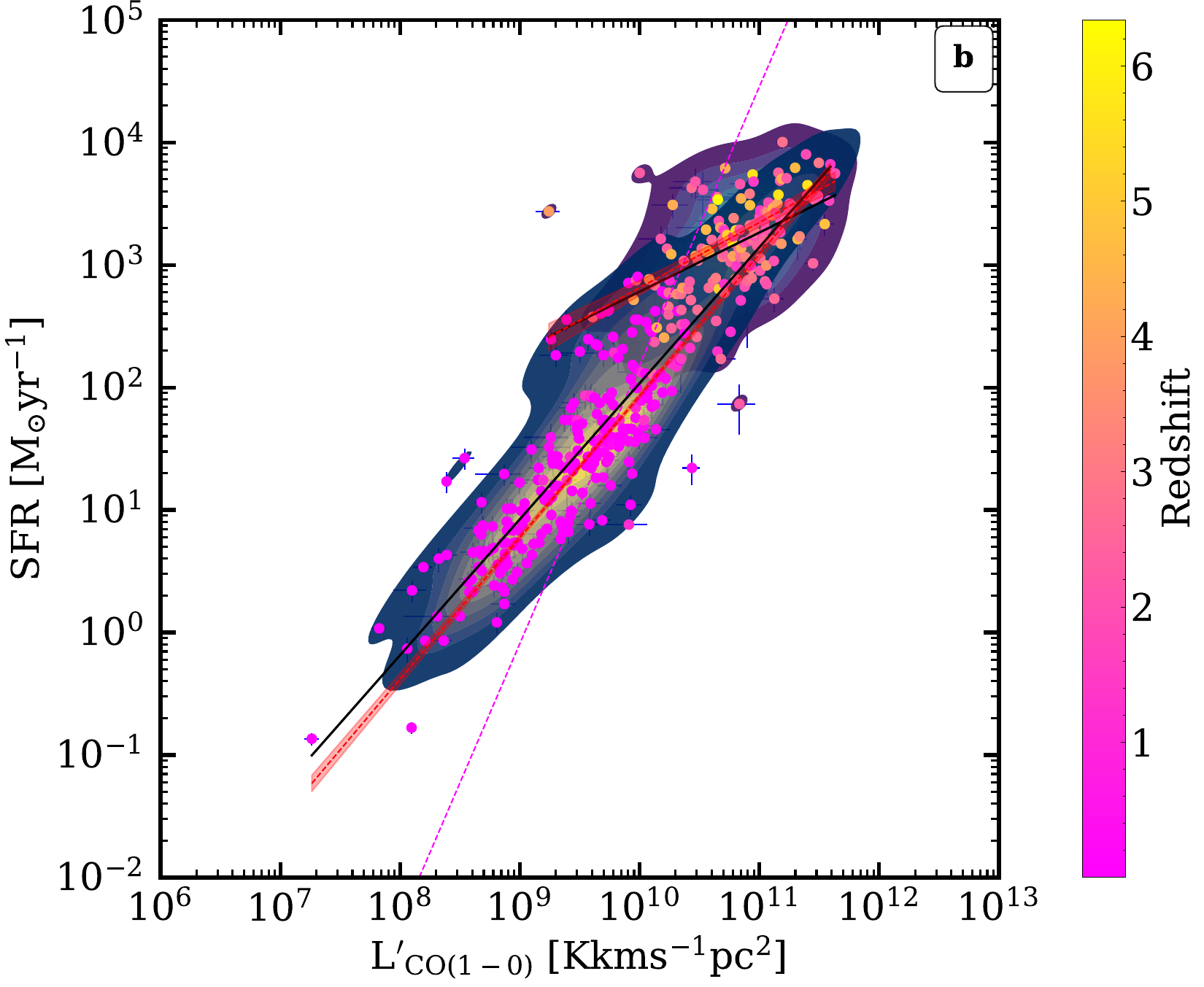}
\caption{SFR versus $\rm L^{'}_{line}$ relations color-coded with respect to redshift.}
\label{fig:SFR_Lliine_z}
\end{figure}

Figure \ref{fig:SFR_Lliine_z} presents the relations between SFR versus $\rm  L^{'}_{[CI](1-0)}$ or $\rm L^{'}_{CO(1-0)}$, color-coded now by redshift. We have performed a separation of the sample with respect to redshift, at \textit{z} = 2, and fit the two samples independently (sample with \textit{z} $>$ 2, and sample with \textit{z} $<$ 2). We derive four relations (for each panel of Fig. \ref{fig:SFR_Lliine_z}) for each tracer (see Table \ref{tab:slopes_with_z} for the derived slopes and intercepts). Both plots clearly show the need for a large dynamic range in redshift, to retrieve the correct global slope, for both species. The clear redshift separation in Fig. \ref{fig:SFR_Lliine_z} states that samples chosen based on a specific redshift range will always have a shallower slope than the global. Quantities with stronger selection biases are consistent with this trend toward shallower slopes. A better retrieval of the true correlation can be achieved by reducing potential selection biases by choosing a combination of samples with significantly different \textit{z}.

\begin{table}[!h]
\centering
\caption{Slope and intercept values for the SFR  versus $\rm L^{'}_{line}$ relations.}
\label{tab:slopes_with_z}
\begin{tabular}{@{}ccccc@{}}
\toprule
\toprule
Model  & \multicolumn{2}{c}{$\rm L^{'}_{[CI]((1-0)}$}           & \multicolumn{2}{c}{$\rm L^{'}_{CO((1-0)}$}           \\ \midrule
                        \multicolumn{5}{c}{\textit{z} $>$ 2 sample}           \\ \midrule
-      & slope ($\beta$) & intercept ($\alpha$)                 & slope ($\beta$) & intercept ($\alpha$)               \\
lmfit  & 0.62 $\pm$ 0.07 & \mysecondcolor{-3.16} $\pm$ 0.75   & 0.53 $\pm$ 0.07 & -2.51 $\pm$ 0.76 \\
linmix & 0.63 $\pm$ 0.08 & \mysecondcolor{-3.22} $\pm$ 0.82   & 0.48 $\pm$ 0.08 & -2.04 $\pm$ 0.85 \\ \midrule
\multicolumn{5}{c}{\textit{z} $<$ 2 sample}           \\ \midrule
-      & slope ($\beta$) & intercept ($\alpha$)                 & slope ($\beta$) & intercept ($\alpha$)               \\
lmfit  & 0.91 $\pm$ 0.03 & -6.54 $\pm$ 0.26   & 1.15 $\pm$ 0.02 & -9.63 $\pm$ 0.26 \\
linmix & 0.99 $\pm$ 0.04 & -7.30 $\pm$ 0.34   & 1.11 $\pm$ 0.03 & -9.07 $\pm$ 0.30  \\
\bottomrule
\end{tabular}
\end{table}

\newpage

\section{Total sample}\label{appendix_total_sample}
Table \ref{tab:table_sample} presents our total sample. The individual numbers of sources with [\ci](1-0), CO(1-0), CO(2-1) and [\cii] are listed along with the total number of sources taken from each sample. 
\begin{table*}[!h]
\centering
\caption{Used sample of sources.}
 \label{tab:table_sample}
 \resizebox{\textwidth}{!}{%
\begin{tabular}{@{}lllclccc@{}}
\toprule
\toprule
\textbf{Reference}                                          & [\ci](1-0)        & CO(1-0)  & CO(2-1) &[\cii]    & Method                       & References & Total sources \\ \midrule
\citet{walter2011survey,alaghband2013using,sharon2016total} & 10 [1]            & 11 [0]   & -       & -        & Interferometric, Single dish & a          & 11 \\
\citet{harrington2016early, harrington2021turbulent}        & 17 [0]            & 17 [0]   & -       & -        & Single dish                  & b          & 17 \\
\citet{montoya2023sensitive}                                & 16 (23) [1]       & 22 (18)  & -       & -        & Interferometric, Single dish & c          & 40 \\
\citet{valentino2020properties}                             & 76 (20) [121]     & 30 (187) & 11 (60) & -        & Interferometric, Single dish & d          & 217 \\
\citet{bothwell2017alma}                                    & 12 [1]            & -        & 9 (4)   & 10 (3)   & Interferometric              & e          & 13 \\
\citet{cormier2015herschel}                                 & -                 & -        & -       & 43 [0]   & Single dish                  & f          & 43 \\
\citet{olsen2017sigame}                                     & -                 & -        & -       & 23 [14]  & Interferometric              & g          & 37 \\
\citet{glazer2024studying}                                  & -                 & -        & -       & 1 [2]    & Interferometric              & h          & 3  \\
\citet{dunne2021dust}                                       & 12 [0]            & 12 [0]   & -       & -        & Interferometric              & i          & 12 \\
\citet{dunne2022dust}                                       & 69 (230) [1]      & 261 (39) & -       & -        & Interferometric, Single dish & j          & 300 \\
\citet{berta2023}                                           & \mysecondcolor{27} (146)          & -        & 36 (137)& -        & Interferometric, Single dish & k          & 172 \\
\citet{castillo2024comparative}     & 18 [2]          & 12 [8]        &  - & -        & Interferometric & l          & 20 \\\midrule
Total                                                       & \mysecondcolor{257}               &  365     & 56      &    77    &                              &            & 885    \\ \bottomrule
\end{tabular}
}
\small
\begin{flushleft}
\textbf{Notes.}\\
Tracer columns represent the number of sources having a detection reported in that transition. The final column represents the total number of sources presented in each original work, including upper limits and/or non-detections. Upper limits are denoted in square brackets, while non-detections are in parentheses. The original studies from which the line flux densities were extracted are listed in the References column.\\
\footnotesize{a: \citet{frayer1999molecular,downes1999detection,guilloteau1999dust,frayer1999molecular,cox2002co,downes2003molecular,genzel2003spatially, neri2003interferometric,sheth2004detection,hainline2004study,kneib2005molecular,greve2005interferometric,tacconi2006high}},  
\footnotesize{b: \citet{borys2006mips,iono2009luminous,sturm2010herschel,stacey2010158,hailey2012herschel,bussmann2013gravitational,geach2015red,canameras2015planck,harrington2016early,khatri2016thermal,yang2017molecular,diaz2017discovery,canameras2017aplanck, canameras2017planck,su2017redshift,iglesias2017near,amodeo2018spectroscopic,canameras2018planck,canameras2018aplanck,geach2018magnified,harrington2018total,dannerbauer2019ultra,frye2019plck,nesvadba2019planck,rivera2019atacama,harrington2019red,berman2022passages, kamieneski2024passages}},  
\footnotesize{c: \citet{veilleux2013fast,spoon2013diagnostics}},  
\footnotesize{d: \citet{walter2011survey,alaghband2013using,liu2015high,kamenetzky2016relations, bothwell2017alma,yang2017molecular,popping2017alma,andreani2018extreme,talia2018alma,canameras2018aplanck,harrington2018total,dannerbauer2019ultra,jin2019discovery,bourne2019relationship,nesvadba2019planck,cortzen2020deceptively}},  
\footnotesize{e: \citet{weiss2013alma}}, 
\footnotesize{f: \citet{madden2013overview,herrera2018shining}}, 
\footnotesize{g: \citet{ouchi2013intensely,kanekar2013search,ota2014alma,gonzalez2014search,willott2015star,maiolino2015assembly,schaerer2015new,capak2015galaxies,inoue2016detection,knudsen2016c,pentericci2016tracing,diaz2017herschel,knudsen2017merger,bradavc2017alma,smit2018rotation}}, 
\footnotesize{h: References refer to the literature sample with 6 $<$ $\textit{z}$ $<$ 7 taken from \citet{glazer2024studying}. 
\citet{maiolino2015assembly,schaerer2015new,watson2015dusty,willott2015star,inoue2016detection,knudsen2016c,pentericci2016tracing,bradavc2017alma,katz2017interpreting,matthee2017alma,bowler2018obscured,hashimoto2018onset,carniani2018kiloparsec,smit2018rotation,hashimoto2019big,matthee2019resolved,bakx2020alma,carniani2020missing,harikane2020large,fujimoto2021alma,ferrara2022alma,molyneux2022spectroscopic,schouws2023alma,sommovigo2022alma,valentino2022archival,wong2022alma,heintz2023gas,fujimoto2024jwst}},  
\footnotesize{i: \citet{pilbratt2010herschel}}, 
\footnotesize{j: \citet{mirabel1990co,tinney1990molecular,young1995fcrao,casoli199612co,zhu1999molecular,curran2000dense,dunne2000scuba,dunne2001scuba,yao2003co,gao2004star,thomas2004distribution,stevens2005dust,chapman2005redshift,chapman2010herschel,weiss2005atomic,weiss2013alma,coppin2006scuba,hainline2006observing,kovacs2006sharc,albrecht2007dust,kuno2007nobeyama,ao2008co,wilson2008luminous,baan2008dense,young2008structure,daddi2009co,chung2009redshift,wu2009continuum,papadopoulos2010cosmic,carilli2010imaging,carilli2011expanded,engel2010most,harris2010co,ivison2010herschel,galametz2011probing,koda2011co,ivison2011tracing,ivison2013herschel,frayer2010green,frayer2018discovery,riechers2011extended,riechers2013dust,riechers2020vla,walter2011survey,walter2012intense,cox2011gas,danielson2011properties,lestrade2011molecular,mckean2011new,iono2012initial,pappalardo2012herschel,schruba2012low,magnelli2012herschel,papadopoulos2012molecular,garcia2012star,thomson2012vla,alaghband2013using,bothwell2013survey,bothwell2017alma,alatalo2013atlas3d,pereira2013herschel,wong2013carma,bussmann2013gravitational,bussmann2015hermes,emonts2013co,sharon2013very,sharon2016total,cooray2014hermes,messias2014herschel,messias2019molecular,negrello2014herschel,negrello2016herschel,swinbank2014alma,tan2014dust,ueda2014cold,liu2015high,rosenberg2015herschel,canameras2015planck,dye2015revealing,aravena2016survey,scoville2016ism,spilker2016alma,alatalo2016shocked,chu2017great,villanueva2017vales,hughes2017vales,Jiao_2017,huynh2017less,oteo2017high,oteo2018extreme,bolatto2017edge,cao2017carma,popping2017alma,Lu_2017,yamashita2017cold,falgarone2017large,wong2017alma,yang2017molecular,yang2019co,clark2018dustpedia,bethermin2018dense,enia2018herschel,pavesi2018hidden,pavesi2018co,perna2018molecular,valentino2018survey,valentino2020properties,wang2018revealing,dannerbauer2019ultra,gomez2019confirming,jiao2019resolved,jiao2021carbon,herrero2019molecular,bourne2019relationship,hunt2019comprehensive,lapham2019spire,sorai2019co,jin2019discovery,kaasinen2019molecular,leung2019ism,nesvadba2019planck,bakx2020alma,boogaard2020alma,michiyama2020discovery,izumi2020alma,berta2021close,ciesla2020hyper,drew2020three,neri2020noema,harrington2021turbulent,dunne2021dust}},  
\footnotesize{k: \citet{nayyeri2016candidate,bakx2018herschel,neri2020noema,cox2023z,ismail2023z}}, 
\footnotesize{l: \citet{serjeant2017scuba,stach2018alma,hill2018high,simpson2019east,simpson2020alma,birkin2021alma,chen2022alma,castillo2023vla,liao2024alma}}  
\end{flushleft}
\end{table*}

\end{appendix}

\end{document}